\documentclass[12pt,epsf,qsymbols]{article}
\pdfoutput=1
\usepackage[utf8]{inputenc}
\usepackage[normalem]{ulem}
\usepackage[english]{babel}
\usepackage{tabularx}
\usepackage{array}
\usepackage{graphics}
\usepackage[pdftex]{graphicx}
\usepackage{psfrag}
\usepackage{epsfig}
\usepackage{subfig}
\usepackage{mathtools}
\usepackage{amssymb}
\usepackage{bbold}
\usepackage{setspace}
\usepackage{rotating}
\usepackage{colortbl}
\usepackage{longtable}
\usepackage{slashed}
\usepackage{braket}
\usepackage{lineno}
\makeatletter
\usepackage{textcomp}
\usepackage[usenames,dvipsnames]{xcolor}
\usepackage{relsize}
\usepackage{cite}

\usepackage{float}

\usepackage{bbm}
\usepackage{bm}
\usepackage{enumerate}
\usepackage{amsmath}
\usepackage{verbatim}

\usepackage{hyperref}
\hypersetup{colorlinks,citecolor=nicegreen,linkcolor=niceblue}
\hypersetup{colorlinks=true}

\usepackage{pifont}
\newcommand{\xmark}{\ding{55}}%

\allowdisplaybreaks

\setlongtables

\newcommand{\bea}{\begin{eqnarray}}
\newcommand{\eea}{\end{eqnarray}}
\newcommand{\beq}{\begin{equation}}
{
\newcommand{\eeq}{\end{equation}}
\newcommand{\ec}{\end{center}}
\newcommand{\bc}{\begin{center}}

\newcommand{\tev}{{\rm TeV}}
\newcommand{\gev}{{\rm GeV}}

\newcommand{\pdir}{p\kern -5.2pt\raise 0.2ex\hbox {/}}

\newcommand{\vdir}{v\kern -5.75pt\raise 0.15ex\hbox {/}}
\newcommand{\kdir}{k\kern -5.75pt\raise 0.15ex\hbox {/}}
\newcommand{\epsdir}{\epsilon\kern -5.0pt\raise 0.15ex\hbox {/}}
\newcommand{\bvdir}{\bar{v}\kern -5.75pt\raise 0.15ex\hbox {/}}
\newcommand{\Ddir}{D\kern -7.75pt\raise 0.20ex\hbox {/}}
\newcommand{\Adir}{A\kern -7.75pt\raise 0.20ex\hbox {/}}
\newcommand{\ldir}{l\kern -5.0pt\raise 0.2ex\hbox{/}}
\newcommand{\varepsdir}{\varepsilon\kern -5.5pt\raise 0.15ex\hbox{/}}

\newcommand{\nn}{\nonumber}

\newcommand{\hadtau}{\tau_{\text{had}}}
\newcommand{\lumi}{{\mathcal{L}}_{\text{int}}}
\newcommand{\invfb}{fb$^{-1}$ }

\makeatother

\definecolor{niceblue}{rgb}{0.15,0.15,0.6}
\definecolor{nicegreen}{rgb}{0.1,0.5,0.1}
\definecolor{Red}{rgb}{1.,0.,0.}

\definecolor{Green}{rgb}{0.2,.7,0.2}

\begin{document}
\unitlength = 1mm

\thispagestyle{empty} 
\begin{flushright}
\begin{tabular}{l}
{\tt \footnotesize LPT-Orsay-18-81}\\
\end{tabular}
\end{flushright}
\begin{center}
\vskip 3.4cm\par
{\par\centering \textbf{\LARGE  
\Large \bf Closing the window on single leptoquark solutions  \\[0.3em] to the $B$-physics anomalies}}
\vskip 1.2cm\par
{\scalebox{.85}{\par\centering \large  
\sc A.~Angelescu$^a$, D.~Be\v{c}irevi\'c$^b$, D.A.~Faroughy$^c$ and O.~Sumensari$^{d,e}$}
{\par\centering \vskip 0.7 cm\par}
{\sl 
$^a$~{Department of Physics and Astronomy\\ University of Nebraska-Lincoln, Lincoln, NE, 68588, USA.}}\\
{\par\centering \vskip 0.25 cm\par}
{\sl 
$^b$~Laboratoire de Physique Th\'eorique (B\^at.~210)\\
CNRS, Univ. Paris-Sud, Universit\'e Paris-Saclay, 91405 Orsay, France.}\\
{\par\centering \vskip 0.25 cm\par}
{\sl 
$^c$~J.~Stefan Institute, Jamova 39, P.~O.~Box 3000, 1001 Ljubljana, Slovenia.}
\\
{\par\centering \vskip 0.25 cm\par}
{\sl 
$^d$~Istituto Nazionale Fisica Nucleare, Sezione di Padova, I-35131 Padova, Italy}\\
{\par\centering \vskip 0.25 cm\par}
{\sl 
$^e$~Dipartamento di Fisica e Astronomia ``G.~Galilei", Università di Padova, Italy}\\

{\vskip 1.65cm\par}}
\end{center}

\vskip 0.85cm
\begin{abstract}
We examine various scenarios in which the Standard Model is extended by a light leptoquark state to solve for one or both $B$-physics anomalies,  viz. $R_{D^{(\ast)}}^\mathrm{exp}> R_{D^{(\ast)}}^\mathrm{SM}$ or/and $R_{K^{(\ast)}}^\mathrm{exp}< R_{K^{(\ast)}}^\mathrm{SM}$. To do so we combine the constraints arising both from the low-energy observables and from direct searches at the LHC. We find that none of the scalar leptoquarks of mass $m_\mathrm{LQ} \simeq 1$~TeV can alone accommodate the above mentioned anomalies. The only single leptoquark scenario which can provide a viable solution for $m_\mathrm{LQ} \simeq 1\div 2$~TeV is a vector leptoquark, known as $U_1$, which we re-examine in its minimal form (letting only left-handed couplings to have non-zero values). We find that the limits deduced from direct searches are complementary to the low-energy physics constraints. In particular, we find a rather stable lower bound on the lepton flavor violating $b\to s\ell_1^\pm\ell_2^\mp$ modes, such as $\mathcal{B}(B\to K\mu\tau)$. Improving the experimental upper bound on $\mathcal{B}(B\to K\mu\tau)$ by two orders of magnitude could compromise the viability of the minimal $U_1$ model as well. 
\end{abstract}
\newpage
\setcounter{page}{1}
\setcounter{footnote}{0}
\setcounter{equation}{0}
\noindent

\renewcommand{\thefootnote}{\arabic{footnote}}

\setcounter{footnote}{0}

\tableofcontents

\newpage


\section{Introduction}

Over the past several years we witnessed a growing interest in theoretical studies of the origin of lepton flavor universality violation (LFUV), motivated by a number of experimental hints 
in weak decays of $B$-mesons pointing towards LFUV. The first such indication was reported by BaBar in Refs.~\cite{Lees:2012xj,Lees:2013uzd} in which they measured 
\begin{equation}
R_{D^{(\ast)}} = \left. \dfrac{\mathcal{B}(B\to D^{(\ast)} \tau\bar{\nu})}{\mathcal{B}(B\to D^{(\ast)} l \bar{\nu})}\right|_{l\in \{e,\mu\}},
\label{eq:RD_definition}
\end{equation}
and found an excess in $\mathcal{B}(B\to D^{(\ast)} \tau\bar{\nu})$. Since that time, Belle and LHCb measured the same ratio~\cite{Huschle:2015rga,Aaij:2015yra,Hirose:2016wfn,Sato:2016svk,Abdesselam:2016cgx} and observed a similar feature, namely that the measured 
$R_{D^{(\ast)}}^\mathrm{exp}$ is {\it larger} than $R_{D^{(\ast)}}^\mathrm{SM}$, the value predicted in the Standard Model (SM). 
The most recent HFLAV averages are~\cite{Amhis:2016xyh}:
\bea\label{eq:RDstar}
R_{D} =0.41 (5) \,,\qquad 
R_{D^{\ast }} = 0.31(2)\,,
\eea 
which, when combined, give $3.8\,\sigma$ excess with respect to (w.r.t.) the SM values, $R_{D}^\mathrm{SM}=0.300(8)$~\cite{Lattice:2015rga,Na:2015kha,Aoki:2016frl}, and $R_{D^{\ast}}^\mathrm{SM}=0.257(3)$~\cite{Bigi:2017jbd,Bernlochner:2017jka}. Apart from the reduction of a significant part of the systematic experimental errors, the advantage of considering the ratio of decay rates lies in the fact that the Cabibbo--Kobayashi-Maskawa (CKM) factors cancel out and in the fact that the sensitivity to hadronic uncertainties is much smaller than
 it is in the case with one of the branching fractions alone $\mathcal{B}(B\to D^{(\ast)} \ell\bar{\nu})$ 
, $\ell\in \{e,\mu, \tau\}$. Even though a $5\, \sigma$ significance of LFUV in the tree-level $b\to c\ell\bar\nu$ decay has not yet been reached, the experimentalists of LHCb were able to confirm the same tendency in another hadronic environment. They measured~\cite{Aaij:2017tyk} 
\bea
R_{J/\psi} =   \dfrac{\mathcal{B}(B_c\to J/\psi \tau\bar{\nu})}{\mathcal{B}(B_c\to J/\psi \mu \bar{\nu})} = 0.71\pm 0.25\,,
\eea
which again appears to be $\approx 2\,\sigma$ larger than its SM value. 

Another indication of the LFUV came from the weak decays mediated by a flavor changing neutral current (FCNC), $b\to sl^+l^-$. The experimentalists of LHCb measured 
\begin{equation}
R_{K^{(\ast)}}^{[q_1^2, q_2^2]} =  \dfrac{\mathcal{B}'(B\to K^{(\ast)} \mu\mu)}{\mathcal{B}'(B\to K^{(\ast)} ee)}  \,,
\label{eq:RK_definition}
\end{equation}
where $\mathcal{B}'$ stands for the partial branching fraction comprising $q^2=(p_{l^+}+p_{l^-})^2$  between $q_1^2$ and $q_2^2$ (in units of $\gev^2$). They reported~\cite{Aaij:2014ora,Aaij:2017vbb}:
\bea
R_K\equiv R_{K^{+}}^{[1,6]}=0.75\pm 0.09, \quad R_{K^\ast}\equiv R_{K^{\ast 0}}^{[1.1,6]}=0.71\pm 0.10,\quad R_{K^{\ast 0}}^{[0.045,1.1]}=0.68\pm 0.10,
\eea
which are $\approx 2.5\,\sigma$ {\it smaller} than the values predicted in the SM~\cite{Bordone:2016gaq}. Although the experimental confirmation of these results is still lacking and the further improvement is needed to increase the significance of the observed deviations w.r.t. the SM, the fact that the indications of LFUV do not concern only the tree-level decays but also those that are in the SM generated by quantum loops, stimulated a considerable activity in the flavor physics community. The observations that $R_{D^{(\ast)}}^\mathrm{exp}> R_{D^{(\ast)}}^\mathrm{SM}$ and $R_{K^{(\ast)}}^\mathrm{exp}< R_{K^{(\ast)}}^\mathrm{SM}$ are commonly referred to as the ``{\it $B$-physics anomalies}".~\footnote{For shortness, we only write $R_{D^{(\ast)}}^\mathrm{exp}> R_{D^{(\ast)}}^\mathrm{SM}$ and $R_{K^{(\ast)}}^\mathrm{exp}< R_{K^{(\ast)}}^\mathrm{SM}$, but one should also keep in mind that $R_{J/\psi}^\mathrm{exp}> R_{J/\psi}^\mathrm{SM}$. }

Apart from the mass effects, different phase space, and moderate hadronic uncertainties, no other reason can be found in the SM to explain the above-mentioned anomalies. In other words, in order to explain (or merely accommodate) the observed deviations w.r.t. SM, one needs to invoke a scenario of physics beyond the SM. 
The simplest effective scenario is to introduce the couplings of left-handed fermions to new vector bosons. In practice that means that the New Physics (NP) effective operators will be of the ``$(V-A)\times (V-A)$" form, which are then fit with the measured $R_{K^{(\ast)},D^{(\ast)}}^\mathrm{exp}$ values to reveal that the NP scale affecting the charged current processes is very different from the one needed to explain the LFUV in the FCNC processes. In such a situation one needs to build a model in which one adjusts Yukawa couplings while keeping the NP scale the same for both types of $B$-physics anomalies. Another possibility is to build a model in which other Lorentz structures are also allowed (such as the scalar and/or tensor currents). One of the most popular scenarios in which most of these ideas can be tested are those based on the introduction of one or more leptoquark states, the colored new bosons which couple to both quarks and leptons. 
In this paper we are going to test the possibility of a single scalar or vector leptoquark (LQ) as a mediator of NP that can accommodate one or both of the $B$-physics anomalies, i.e. $R_{K^{(\ast)}}^\mathrm{exp}< R_{K^{(\ast)}}^\mathrm{SM}$ and/or $R_{D^{(\ast)}}^\mathrm{exp}> R_{D^{(\ast)}}^\mathrm{SM}$.
We will go through various LQ scenarios to examine if any of them remains plausible. In doing so we will go for a minimalistic approach, i.e. that by extending the SM by one LQ involves the least number of new parameters, which then permits to test the model experimentally. In that respect, the current paper is similar to Ref.~\cite{Becirevic:2016oho} and could be viewed as its update. The new element is the fact that we combine the updated constraints arising from numerous low-energy physics observables 
with those deduced from the direct searches at the LHC.~\footnote{Research on combining the low-energy constraints to building a SUSY inspired model with the results of direct searches at the LHC has been  reported in Ref.~\cite{Altmannshofer:2017poe}.} Since this kind of models can give rise to lepton flavor violation (LFV)~\cite{Glashow:2014iga},  
a particular attention will be devoted to the vector LQ model ($U_1$) for which we will show that the improved experimental bounds on $\mathcal{B}(B_s\to \mu\tau)$ and/or $\mathcal{B}(B\to K^{(\ast )}\mu\tau)$, could validate or discard the model in its minimalistic form. 
Before we embark on the details of this work, we should emphasize that by a single LQ we mean one multiplet (singlet, doublet or triplet) of mass degenerate LQ's which carry the same quantum numbers of the SM gauge group.

This paper is organized as follows. In Sec.~\ref{sec:eft}, we remind the reader of a low-energy effective description of $b \to s \ell \ell$ and $b \to c \ell \bar\nu$ transitions, and we present the $1\,\sigma$ bounds on the effective coefficients which accompany the hadronic matrix elements of dimension-six operators relevant to a general NP scenario. 
In Sec.~\ref{sec:lq-list}, we briefly go through the list of the single LQ solutions to the $B$-physics anomalies, and for each of them we compute the effective Wilson coefficients relevant to $R_{K^{(\ast)}}$ and to $R_{D^{(\ast)}}$, respectively. In Sec.~\ref{sec:LHC}, we discuss the bounds on the Yukawa couplings arising from the direct searches at LHC. Besides LQ pair production, we also comment on the measurements of tails of high-$p_T$ distributions of lepton pairs (in $pp\to \mu\mu, \tau\tau$ decays) which can be modified w.r.t. the SM by the $t$-channel LQ exchange. In Sec.~\ref{sec:which-LQ} the low-energy and high energy constraints are combined. We will then show that the only one that can accommodate both anomalies and survive all the constraints turns out to be the vector LQ $U_1$, the phenomenology of which is analyzed in more detail in Sec.~\ref{sec:results}, where we also show that the LFV processes $b\to s\mu\tau$, together with high-$p_T$ lepton tails at LHC, can provide complementary constraints and can be used to validate or invalidate model in its minimalistic form. We summarize our findings in Sec.~\ref{sec:conclusions}.

\

\section{Effective Theory Description}
\label{sec:eft}
Effective theories provide an efficient way to describe the low-energy physics processes in which the short-distance physics is encoded in the so called Wilson coefficients, which are computed perturbatively, 
whereas the remaining long-distance physics is expressed in terms of a number of effective (dimension-six) operators, higher dimension operators being suppressed by powers of a high energy scale. 
In the SM the matching between the full and effective theories for the loop-induced $b\to s\ell^+\ell^-$ process includes the next-to-next-to-leading logarithmic corrections and the resummation of potentially large logarithms has been made by means of the renormalization group equations~\cite{Bobeth:1999mk}. Looking for NP in this context means to look for the non-SM contributions to the Wilson coefficients corresponding to the operators already present in the SM, in addition to those that are 
not present to the SM but which are allowed on the Lorentz symmetry grounds. In this Section, for definiteness, we remind the reader of the low-energy effective theories relevant to both $b\to s\ell^+\ell^-$ and $b\to c\ell \bar \nu$ decays. 
\subsection{$b\to s\ell_1^-\ell_2^+$ and $R_{K^{(\ast)}}$}
\label{ssec:RK}

Since we will be concerned with both lepton flavor conserving and LFV decay modes we will describe the effective Hamiltonian for a generic $b\to s \ell_1^- \ell_2^+$, with $\ell_{1,2}\in\{e,\mu,\tau\}$, which can be written as

\begin{equation}
\label{eq:hamiltonian-bsll}
\begin{split}
  \mathcal{H}_{\mathrm{eff}} = -\frac{4
    G_F}{\sqrt{2}}V_{tb}V_{ts}^* &\Bigg{\lbrace} \sum_{i=1}^6
  C_i(\mu)\mathcal{O}_i(\mu)+\sum_{i=7,8}
  \Big{[}C_i(\mu)\mathcal{O}_i(\mu)+\left(C_{i}(\mu)\right)^\prime \left(\mathcal{O}_{i}(\mu)\right)^\prime\Big{]}\\
& + \sum_{i=9,10,S,P}
  \Big{[} C^{\ell_1 \ell_2}_i(\mu)\mathcal{O}^{\ell_1 \ell_2}_i(\mu) + \left(C^{\ell_1 \ell_2}_{i}(\mu)\right)^\prime \left(\mathcal{O}^{\ell_1 \ell_2}_{i}(\mu)\right)^\prime\Big{]}\Bigg{\rbrace}
+\mathrm{h.c.}
\end{split}
\end{equation}
$C_i(\mu)$ and $C_i^{\ell_1 \ell_2}(\mu)$ are the Wilson coefficients, while the effective operators relevant to our study are defined by
\begin{align}
\label{eq:C_LFV}
\begin{split}
\mathcal{O}_{9}^{\ell_1\ell_2}
  &=\frac{e^2}{(4\pi)^2}(\bar{s}\gamma_\mu P_{L}
    b)(\bar{\ell}_1\gamma^\mu\ell_2)\,, \qquad\qquad\hspace*{0.4cm}
\mathcal{O}_{S}^{\ell_1\ell_2} =
  \frac{e^2}{(4\pi)^2}(\bar{s} P_{R} b)(\bar{\ell}_1 \ell_2)\,,\\
    \mathcal{O}_{10}^{\ell_1\ell_2} &=
    \frac{e^2}{(4\pi)^2}(\bar{s}\gamma_\mu P_{L}
    b)(\bar{\ell}_1\gamma^\mu\gamma^5\ell_2)\,,\qquad\qquad
\mathcal{O}_{P}^{\ell_1\ell_2} =
  \frac{e^2}{(4\pi)^2}(\bar{s} P_{R} b)(\bar{\ell}_1 \gamma^5 \ell_2)\,,\\
\end{split}
\end{align}
in addition to the electromagnetic penguin operator,
$\mathcal{O}_7=\left( e/(4\pi)^2\right) m_b (\bar{s}\sigma_{\mu\nu}P_R
b)F^{\mu\nu}$. The primed quantities in Eq.~\eqref{eq:hamiltonian-bsll} correspond to the chirality flipped operators, $\mathcal{O}_i^\prime$, which are obtained from $\mathcal{O}_i$ after replacing $P_L\leftrightarrow P_R$. From this Hamiltonian it is straightforward to compute the decay rates for $B_s\to \ell_1 \ell_2$ and $B\to K^{(\ast)} \ell_1 \ell_2$, cf.~Ref.~\cite{Becirevic:2016zri}. In the following we will omit the dependence on the renormalization scale and take $C_i \equiv C_i(m_b)$. 

\begin{figure}[t!]
  \centering
  \includegraphics[width=0.6\textwidth]{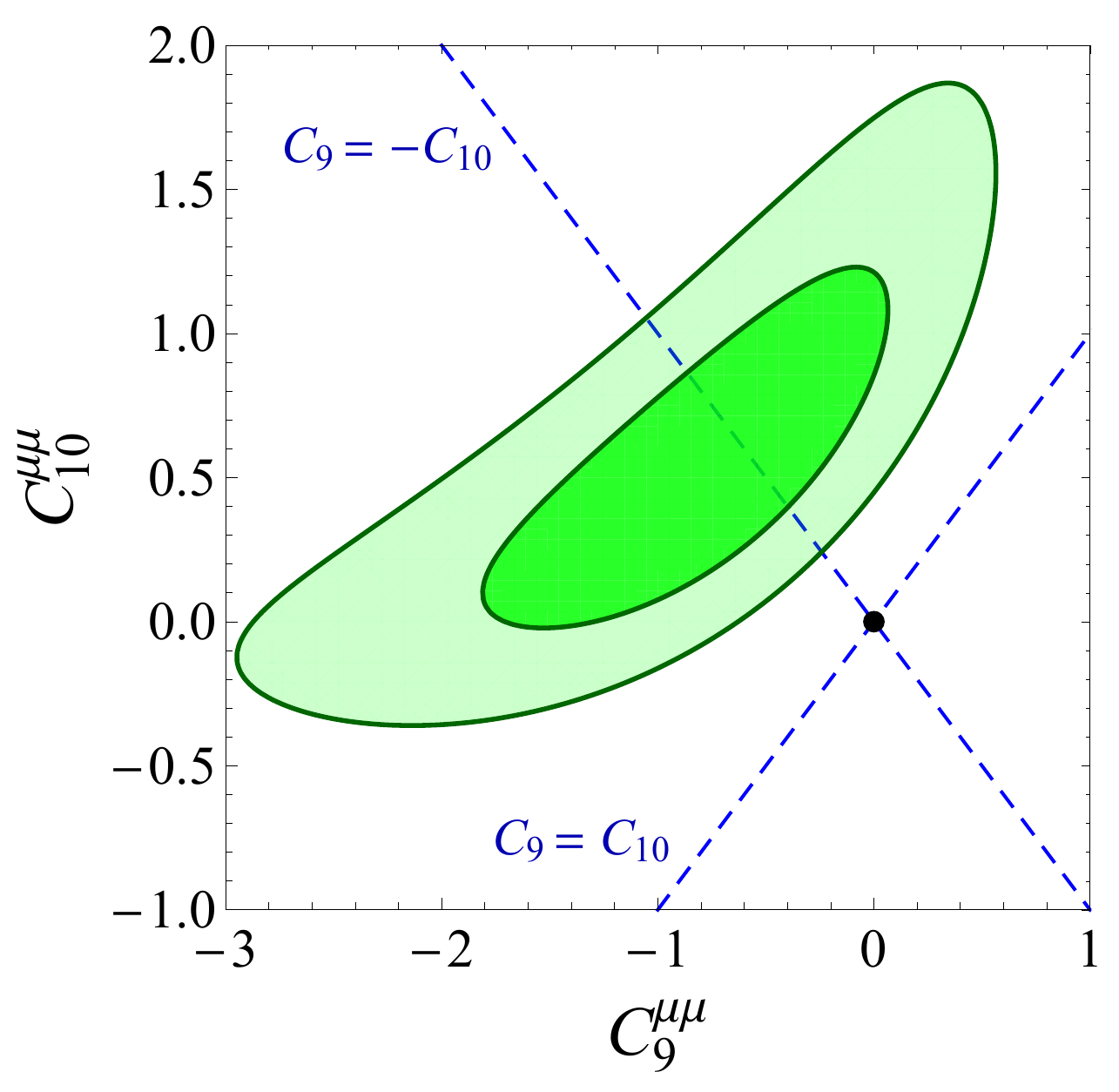}
  \caption{ \sl \small Low-energy fit to $R_K$, $R_{K^\ast}$ and $\mathcal{B}(B_s\to\mu\mu)$ in the plane $C_9^{\mu\mu}$ vs.~$C_{10}^{\mu\mu}$. Darker (lighter) region is allowed to $1\, \sigma$ ($2\,\sigma$) accuracy. Blue dashed lines correspond to scenarios with $C_9^{\mu\mu}=\pm C_{10}^{\mu\mu}$, while the black dot denotes the SM point.}
  \label{fig:fits-RK}
\end{figure}
By assuming that the NP couplings to electrons are negligible, it has been established that $R_K$ and $R_{K^\ast}$ can be explained by a purely vector Wilson coefficient, $C_9^{\mu\mu}<0$, or by a left-handed combination, $C_9^{\mu\mu}=-C_{10}^{\mu\mu}<0$.~\footnote{An explanation of $R_{K^{(\ast)}}$ by NP couplings to electrons is disfavored by global analysis of the $b\to s\mu\mu$ observables, cf.~Ref.~\cite{global}.}  
 The result of our fit, illustrated in Fig.~\ref{fig:fits-RK} is

\begin{equation}
\label{eq:C9-exp}
C_9^{\mu\mu}=-C_{10}^{\mu\mu} \in (-0.85,-0.50)\,,
\end{equation}
\noindent which deviates from the SM by almost $4\sigma$. In this fit we used  $R_{K^{(\ast)}}^{\mathrm{exp}}$~\cite{Aaij:2014ora,Aaij:2017vbb}, and the theoretically clean $\mathcal{B}(B_s\to \mu\mu)^{\mathrm{exp}}=\left(3.0\pm 0.6^{+0.3}_{-0.2}\right) \times 10^{-9}$~\cite{Aaij:2017vad}.~\footnote{Notice that the measured $\mathcal{B}(B_s\to \mu\mu)^{\mathrm{exp}}$, agrees with the SM value $\mathcal{B}(B_s\to~\mu\mu)^{\mathrm{SM}}=\left(3.65\pm 0.23\right)\times 10^{-9}$~\cite{Bobeth:2013uxa}.}

The possibility of having $C_9^{\mu\mu}=-C_{10}^{\mu\mu}$ is particularly interesting because it is realized in several LQ scenarios~\cite{Dorsner:2016wpm}.
From now on, for notational simplicity, we will omit the ``$\mu\mu$"-superscript.

\subsection{$b\to c\ell\bar \nu$ and $R_{D^{(\ast)}}$}
\label{ssec:RD}

The most general low-energy effective Lagrangian involving all of the dimension-six operators capable to generate a (semi-)leptonic decay via charged currents reads
	\begin{align}
		\label{eq:lagrangian-lep-semilep}
		\mathcal{L}_{\mathrm{eff}} &= -2\sqrt{2}G_F V_{u d} \Big{[}(1+g_{V_L})\,(\overline{u}_{L}\gamma_\mu {d}_{L}) (\overline{\ell}_L\gamma^\mu\nu_L) + g_{V_R}\,(\overline{u}_{R}\gamma_\mu {d}_{R}) (\overline{\ell}_L\gamma^\mu\nu_L) \\[0.3em]
		&+g_{S_L}(\mu)\,(\overline{u}_{R} d_{L})(\overline{\ell}_R \nu_L)+g_{S_R}(\mu)\,(\overline{u}_{L} d_{R})(\overline{\ell}_R \nu_L)+g_T(\mu)\,(\overline{u}_R \sigma_{\mu\nu}d_L)(\overline{\ell}_R \sigma^{\mu\nu} \nu_L)\Big{]}+\mathrm{h.c.}\,,\nonumber
	\end{align}

\noindent where $u$ and $d$ stand for a generic up- and down-type quarks, and $g_{i}\equiv g_i(m_b)$ are the effective NP couplings with $i \in \lbrace V_{L(R)},S_{L(R)},T \rbrace$. 
In order to describe the anomalies observed in the exclusive $b\to c\ell\bar \nu$ decays one necessarily needs to introduce the new bosonic fields above the electroweak scale. Such an extended theory should also respect the $SU(2)_L \times U(1)_Y$ symmetry which means that $g_{V_R}$ should be lepton flavor universal at dimension-six, and as such it is irrelevant for the discussion that follows~\cite{Aebischer:2015fzz,Jung:2018lfu}. 
In other words, we are left with four effective coefficients, $g_{V_L}$, $g_{S_L}$, $g_{S_R}$ and $g_T$ which can potentially contribute to $R_{D^{(\ast)}}$.

To determine the allowed values of the effective couplings $g_{i}$ we assume that NP only contributes to the transition $b\to c\tau \bar{\nu}$, and that its effect is negligible to the electron and muon modes.~\footnote{That assumption is a very good approximation to the realistic situation. As we shall see, we find that the couplings of leptoquarks to $b$ and $\tau$ are indeed much larger than those involving muons so that the physics discussion of $R_{D^{(\ast )}}$ remains unchanged after setting the couplings to muons to zero.} We use the $B\to D$ semileptonic form factors computed by means of lattice QCD in Refs.~\cite{Lattice:2015rga,Na:2015kha}. Since the $B\to D^\ast$ form factors at non-zero recoil are still not available from LQCD, we consider the ones extracted from the measured angular distribution of $B\to D^\ast (\to D\pi)l\bar \nu$ ($l\in \{e,\mu\}$) given in Ref.~\cite{Amhis:2016xyh}, and combine them with the ratios $A_0(q^2)/A_1(q^2)$ and $T_{1-3}(q^2)/A_1(q^2)$ computed in Ref.~\cite{Bernlochner:2017jka}. By using these theoretical inputs and the experimental values given in Eq.~\eqref{eq:RDstar} we were able to constrain the values of effective coefficients in Eq.~\eqref{eq:lagrangian-lep-semilep} and thus accommodate $R_{D^{(\ast)}}$. The first solution we consider is the coefficient $g_{V_L}>0$, which corresponds to an overall rescaling 
of the SM. The allowed  $1\,\sigma$ range in this case reads~\footnote{We disregard the solution with large and negative values of $g_{V_L}$, since this possibility would require excessively large NP couplings.}

\begin{equation}
\label{eq:RDfit}
g_{V_L} \Big{\vert}_{b\to c \tau \nu_{\tau}} \in (0.09,0.13)\,.
\end{equation}

\noindent Other solutions involving the coefficients $g_{S_L}$ and/or $g_T$ have also been considered in the literature~\cite{Freytsis:2015qca,Becirevic:2018uab,Becirevic:2012jf}. In particular, two specific scalar LQ scenarios predict $g_{S_L} = \pm 4\, g_T$ at the scale $\mu=m_\mathrm{LQ}$, which is of the order of $1$~TeV~\cite{Dorsner:2016wpm}. That relation is modified when running down to $\mu=m_b$. After including in the renormalization group running the one-loop electroweak corrections, in addition to the three-loop QCD anomalous dimensions, the relations $g_{S_L} = + 4\, g_T$ and $g_{S_L} = - 4\, g_T$ become  $g_{S_L}\approx +  8.14\, g_T$  and $g_{S_L} \approx -8.5 \, g_T$ at $\mu=m_b$, respectively~\cite{Gonzalez-Alonso:2017iyc}. 
%
\begin{figure}[t!]
  \centering
  \includegraphics[width=0.5\textwidth]{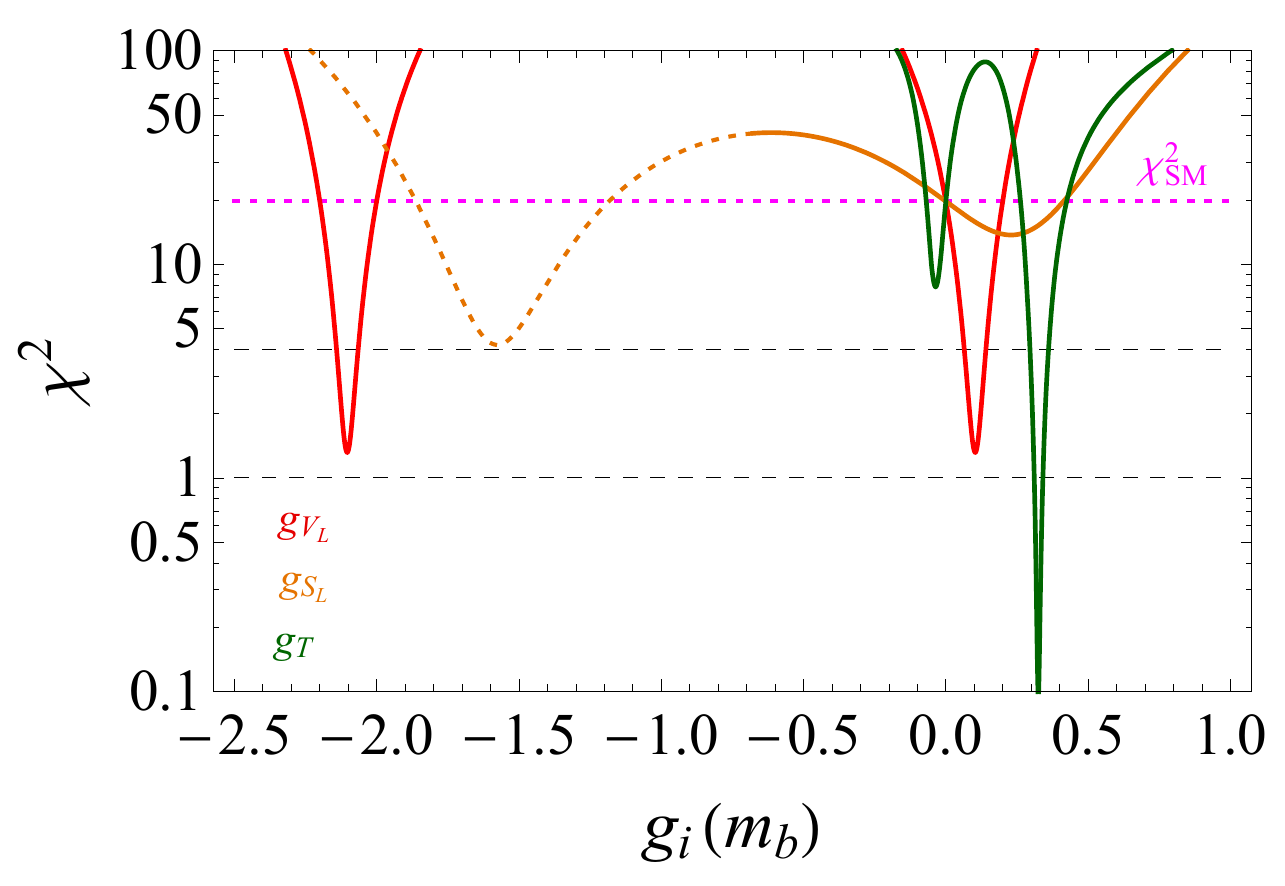}~  \includegraphics[width=0.5\textwidth]{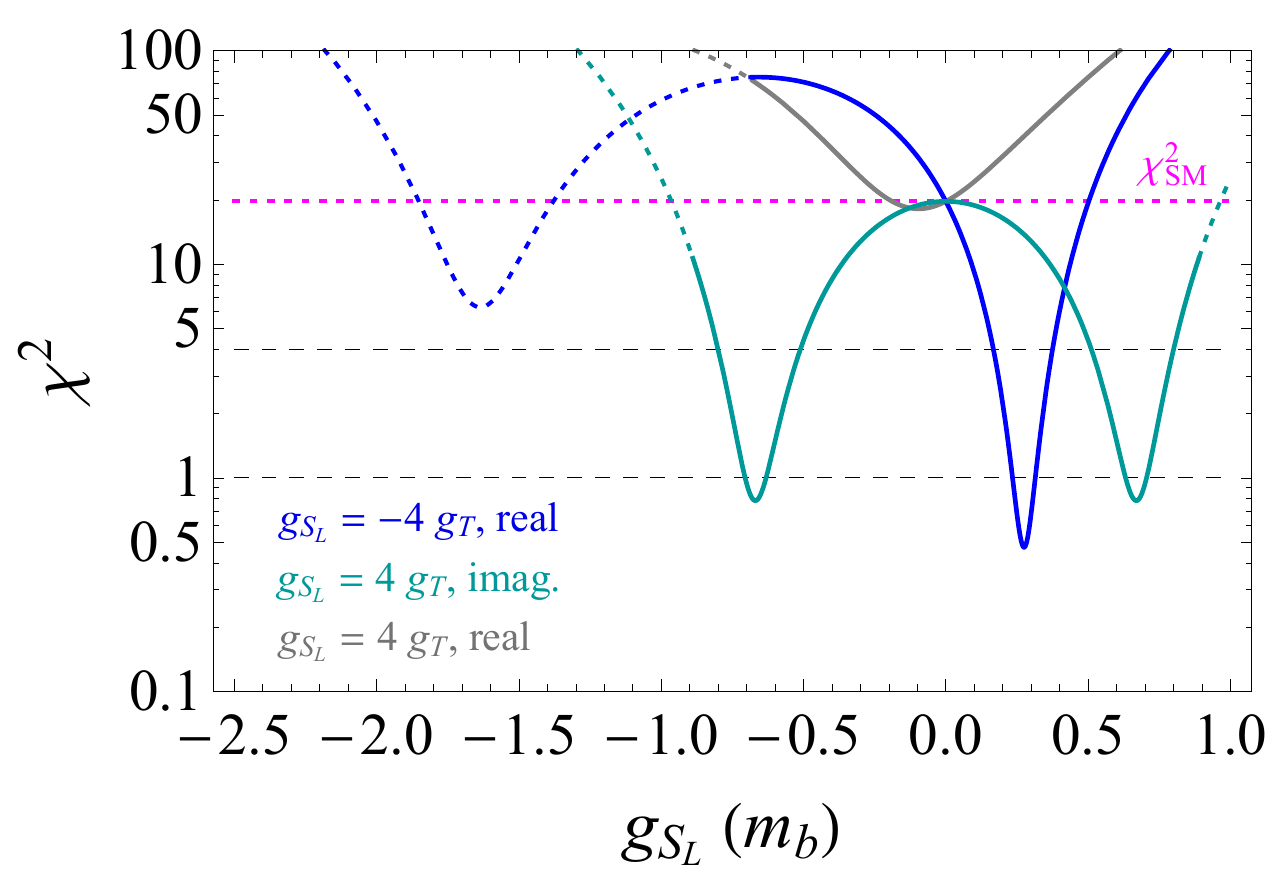}
  \caption{ \sl \small $\chi^2$ values for individual effective coefficients fits of $R_D$ and $R_{D^\ast}$, compared to the SM value, $\chi^2_{\mathrm{SM}}\approx 19.7$. In the left panel, $\chi^2$ is plotted against  $g_{V_L}$, $g_{S_L}$ and $g_T$ at $\mu=m_b$. In the right panel, $\chi^2$ is plotted against  $g_{S_L}(m_b)$ by assuming $g_{S_L} = \pm 4\, g_T$ at $\mu=1~\mathrm{TeV}$, for purely imaginary and real couplings. The dashed regions correspond to the values excluded by the $B_c$-lifetime constraints, see text for details.}
  \label{fig:fits-RD}
\end{figure}
The low-energy fits to various combinations of effective coefficients are shown in Fig~\ref{fig:fits-RD}. In the same plot, we superimpose the limits derived from the $B_c$-meson lifetime, which is particularly efficient to constraint the pseudoscalar contribution~\cite{Li:2016vvp,Alonso:2016oyd}. More precisely, we consider the conservative limit $\mathcal{B}(B_c\to \tau \bar{\nu}) \lesssim 30\%$ and the expression
\begin{equation}
\label{eq:leptonicB}
\mathcal{B}(B_c\to \tau \bar{\nu}) = \tau_{B_c} \dfrac{m_{B_c} f_{B_c}^2 G_F^2 |V_{cb}|^2}{8 \pi} m_\tau^2 \left( 1- \dfrac{m_\tau^2}{m_{B_c}^2} \right)^2 \Bigg{|} 1+ g_{V_L} + \dfrac{ \left(g_{S_R}-g_{S_L}\right)m_{B_c}^2}{m_\tau (m_b+m_c)}\Bigg{|}^2\,,
\end{equation}

\noindent from which we derive that

\begin{equation}
\label{eq:gP-Bclifetime}
g_P (m_b) \equiv g_{S_R}(m_b)-g_{S_L}(m_b)\in (-1.14,0.68)\,.
\end{equation}

\noindent By combining this constraint with the low-energy fit to $R_{D^{(\ast)}}$ described above, we conclude in Fig.~\ref{fig:fits-RD} that not only the scenario with $g_{V_L}>0$ can accommodate $R_{D^{(\ast)}}$, but also other scenarios such as $g_T(m_b)\neq 0$, $g_{S_L}= -4\, g_T>0$. Furthermore, plausible solutions are obtained for $g_{S_L}= \pm 4\, g_T$ but allowing the couplings to be mostly imaginary. These findings are in agreement with the literature, cf.~Ref.~\cite{Freytsis:2015qca}, which are updated in this paper with the most recent experimental and theoretical inputs and extended by allowing  the possibility of imaginary couplings.

\section{Leptoquark models for $R_{K^{(\ast)}}$ and/or $R_{D^{(\ast)}}$}
\label{sec:lq-list}

In this Section we briefly review the LQ models proposed to accommodate the $B$-physics anomalies by introducing a single mediator.  
We adopt the notation of Ref.~\cite{Dorsner:2016wpm} and specify the LQ by their SM quantum numbers, $(SU(3)_c,SU(2)_L)_Y$, where the electric charge, $Q=Y+T_3$, is the sum of the hypercharge ($Y$) and the third weak isospin component ($T_3$). In the left-handed doublets, $Q_i=[(V^\dagger u_L)_i~d_{L\,i}]^T$ and $L_i=[(U\nu_L)_i~\ell_{L\,i}]^T$, the matrices $V$ and $U$ are the CKM and Pontecorvo-Maki-Nakagawa-Sakata (PMNS) matrices, respectively. Since the neutrino masses are insignificant for the phenomenology we study in this paper, we can set $U= \mathbb{1}$. 

	\subsection{Scalar leptoquarks}
		\subsubsection{$S_3=(\mathbf{\bar{3}},\mathbf{3})_{1/3}$}
		\label{sec:lq-list-S3}
		
The first scenario we consider is with $S_3=(\mathbf{\bar{3}},\mathbf{3})_{1/3}$, a weak triplet of scalar LQ states with hypercharge $1/3$.~\footnote{We follow a common practice and in this paper we consider the LQ's belonging to the same multiplet to be mass degenerate.} Remarkably, $S_3$ is the only scalar particle that can simultaneously account for $R_K^{\mathrm{exp}}<R_K^{\mathrm{SM}}$ and $R_{K^\ast}^{\mathrm{exp}}<R_{K^\ast}^{\mathrm{SM}}$ at tree-level~\cite{Dorsner:2017ufx,Hiller:2014yaa}. The Yukawa Lagrangian of $S_3$ reads~\cite{Dorsner:2016wpm}

\begin{equation}
\label{eq:S3model}
\mathcal{L}_{S_3}  = y_L^{ij} \, \overline{Q^C_{i}} i \tau_2 ( \tau_k S^k_3) L_{j}+\mathrm{h.c.}\,,
\end{equation}

\noindent where $\tau_k$ is one of the Pauli matrices ($k=1,2,3$), $S_3^k$ is a component of the LQ triplet, and $y_L$ stands for a generic Yukawa matrix. 
We have neglected the LQ couplings to diquarks in Eq.~\eqref{eq:S3model} in order to ensure the proton stability~\cite{Dorsner:2016wpm}. Otherwise one should devise 
an appropriate symmetry to suppress these couplings that are known to be tightly constrained by experimental limits on the proton lifetime.~\footnote{See Ref.~\cite{Dorsner:2017ufx} for illustration of a Grand Unification scenario in which these couplings are absent in a concrete $SU(5)$ set-up.} It is convenient to rewrite Eq.~\eqref{eq:S3model} in terms of charge eigenstates as:
\begin{align}
\begin{split}
\mathcal{L}_{S_3} = &- y_L^{ij} \, \overline{d^C_{L\,i}} \nu_{L\,j}\, S_3^{(1/3)}-\sqrt{2} \, y_L^{ij} \, \overline{d^C_{L\,i}} \ell_{L\,j}\, S_3^{(4/3)}\\[0.4em]
&+\sqrt{2}\,\left(V^\ast y_L\right)_{ij}\, \overline{u^C_{L\,i}} \nu_{L\,j}\, S_3^{(-2/3)}-\left(V^\ast y_L\right)_{ij} \overline{u^C_{L\,i}} \ell_{L\,j}\, S_3^{(1/3)}+\mathrm{h.c.}\,,
\end{split}
\end{align}

\noindent from which one can easily extract the Wilson coefficients for the $b\to s \ell_l^- \ell_k^+$ decay, namely, 
\begin{equation}
\label{eq:C9-S3}
C_9^{kl} = - C_{10}^{kl} = \dfrac{\pi v^2}{ V_{tb}V_{ts}^\ast \alpha_{\mathrm{em}}} \dfrac{y_L^{b k} \left(y_L^{s l}\right)^\ast}{m_{S_3}^2}\,,
\end{equation}

\noindent which is precisely the effective scenario needed to accommodate $R_{K^{(\ast)}}^{\mathrm{exp}}<R_{K^{(\ast)}}^{\mathrm{SM}}$. Notice once again that 
in this discussion $C_{9,10}^{kl}$ refer only to the LQ contribution, that is to be added to the SM value which is non-zero in the lepton flavor conserving case ($l=k$). 
Similarly, we also read off the contribution arising from this LQ to the $b\to c\ell \overline{\nu}_{\ell^\prime}$ transition and find
\begin{align}
\label{eq:gV-S3}
g_{V_L} = -\dfrac{v^2\,y_L^{b\ell^\prime} \left(V y_L^\ast\right)_{c\ell}}{4 V_{cb} \, m_{S_3}^2} = - \dfrac{v^2}{4 m_{S_3}^2} y_{L}^{b\ell^\prime}\Big{[}(y_L^{b\ell})^\ast + \dfrac{V_{cs}}{V_{cb}} (y_L^{s\ell})^\ast+ \dfrac{V_{cd}}{V_{cb}} (y_L^{d\ell})^\ast\Big{]}\,.
\end{align}

\noindent We see that the coupling $y_L^{b\ell}$ gives a negative contribution to $g_{V_L}$, clearly at odds with $R_{D^{(\ast)}}^{\mathrm{exp}}>R_{D^{(\ast)}}^{\mathrm{SM}}$. Note that the terms depending on $y_L^{s\ell}$ and $y_L^{d\ell}$ are not necessarily negative, but they are tightly constrained by other flavor limits, such as $\mathcal{B}(B\to K \bar{\nu}\nu)$ and the frequency of oscillations in the $B_{d,s}^0-\bar B_{d,s}^0$ system ($\Delta m_{B_{d,s}}$), so that this scenario cannot accommodate $R_{D^{(\ast)}}$.

		\subsubsection{$R_2=(\mathbf{3},\mathbf{2})_{7/6}$}
		\label{sec:lq-list-R2}
		
The second scenario we consider is $R_2=(\mathbf{3},\mathbf{2})_{7/6}$, the weak doublet of scalar LQ's with hypercharge $Y=7/6$. 
This scenario is known to be unsatisfactory at tree level because it leads to 
$R_{K^{(\ast)}}>R_{K^{(\ast)}}^{\mathrm{SM}}$, clearly disfavored by the data. One can get around this problem, as proposed in Ref.~\cite{Becirevic:2017jtw}, and generate corrections 
to $R_{K^{(\ast)}}$ at loop-level and accommodate $R_{K^{(\ast)}}^{\mathrm{exp}}<R_{K^{(\ast)}}^{\mathrm{SM}}$. Interestingly, 
$R_2$ is the only scalar LQ for which the proton stability is automatically preserved~\cite{Assad:2017iib}. The most general Lagrangian describing 
the Yukawa interactions with $R_2$ can be written as,
\begin{align}
\label{eq:yuk-R2}
\mathcal{L}_{R_2} = y_R^{ij} \, \overline{Q}_i \ell_{R\,j}\,R_2 - y_L^{ij} \, \overline{u}_{R\,i} {R_2} i \tau_2 L_j + \mathrm{h.c.}\,,
\end{align}

\noindent where $y_L$ and $y_R$ are the Yukawa matrices, and $SU(2)_L$ indices have been omitted for simplicity. More explicitly, in terms of the electric charge eigenstates, 
the Lagrangian~\eqref{eq:yuk-R2} can be written as
\begin{align}
\label{eq:yuk-R2-bis}
\begin{split}
\mathcal{L}_{R_2} &= (V y_R)_{ij} \, \overline{u}_{L\,i} \ell_{R\,j}\,R_2^{(5/3)} + (y_R)_{ij} \, \overline{d}_{L\,i} \ell_{R\,j}\,R_2^{(2/3)}    \\[0.4em]
&+(y_L)_{ij} \bar{u}_{R\,i} \nu_{L\,j}\, R_2^{(2/3)} - (y_L)_{ij} \overline{u}_{R\,i} \ell_{L\,j}\, R_2^{(5/3)} + \mathrm{h.c.}
\end{split}
\end{align}

\noindent As mentioned above, the tree-level contribution to the Wilson coefficients amounts to

\begin{equation}
 C_9^{kl} =  C_{10}^{kl} \stackrel{\mathrm{tree}}{=}  - \dfrac{\pi v^2}{2 V_{tb}V_{ts}^\ast\alpha_{\mathrm{em}}}\dfrac{y_R^{sl} \big{(}y_R^{bk}\big{)}^\ast}{m_{R_2}^2}\,,
\end{equation}

\noindent which leads to $R_{K^{(\ast)}}>R_{K^{(\ast)}}^{\mathrm{SM}}$, in conflict with experiment.  
Instead, if the LQ corrections start at loop-level, which can be achieved by setting $y_R \equiv 0$, one finds~\cite{Becirevic:2017jtw}
\begin{align}\label{eq:C9new}
C_9^{kl}=- C_{10}^{kl} \stackrel{\mathrm{loop}}{=}  \sum_{u,u^\prime \in \{u,c,t\}} {V_{ub} V_{u^\prime s}^\ast\over V_{tb} V_{t s}^\ast } y_L^{u^\prime k} \left( y_L^{u l}\right)^\ast \mathcal{F}(x_u, x_{u^\prime})\,,
\end{align}
where $x_{i}= m_{i}^2/m_W^2$, and the loop function reads, 
\begin{align}
\mathcal{F}(x_u, x_{u^\prime})= {\sqrt{x_u x_{u^\prime}} \over 32\pi\alpha_\mathrm{em}}  &\biggl[ {  x_{u^\prime} (  x_{u^\prime} - 4) \log  x_{u^\prime}\over ( x_{u^\prime}-1) (x_u- x_{u^\prime})( x_{u^\prime}-x_\Delta)}
+ { x_u (  x_u - 4) \log  x_u\over ( x_u-1) (x_{u^\prime} - x_u )( x_u -x_\Delta)} 
\biggr.\nn\\[2.ex]
&\biggl.  - { x_\Delta (x_\Delta -4) \log x_\Delta \over (x_\Delta -1) (x_\Delta - x_u)( x_\Delta - x_{u^\prime} ) } 
\biggr]\,.
\end{align}

\noindent Furthermore, this LQ state contributes to the transition $b\to c \tau \bar{\nu}_{\ell^\prime}$ via the following coefficients,

\begin{equation}
g_{S_L} = 4 \, g_T = \dfrac{v^2}{4 V_{cb}} \dfrac{y_{L}^{c\ell^\prime}\big{(}y_R^{b\ell}\big{)}^\ast}{m_{R_2}^2 } \,,
\end{equation}

\noindent obtained by tree-level matching at the scale $\mu=m_{R_2}$. This scenario can accommodate the observed deviations in $R_{D^{(\ast)}}$ for complex and even purely imaginary couplings, as it can be seen in Fig.~\ref{fig:fits-RD}, and in Refs.~\cite{Becirevic:2018uab,Sakaki:2013bfa,Hiller:2016kry}. 

A simultaneous explanation of $R_{K^{(\ast)}}$ and $R_{D^{(\ast)}}$ is, however, excluded due to the stringent limits on $\mathcal{B}(\tau \to \mu\gamma)$, which would receive a contribution enhanced by a factor of $m_t/m_\tau$ at the amplitude level, as discussed in Ref.~\cite{Becirevic:2017jtw}. Therefore, this scenario can accommodate either $R_{K^{(\ast)}}$ or $R_{D^{(\ast)}}$, but not both.

		\subsubsection{$\widetilde{R}_2=(\mathbf{3},\mathbf{2})_{1/6}$}
	\label{sec:lq-list-R2tilde}
	
Another important scenario to consider is $\widetilde{R}_2=(\mathbf{3},\mathbf{2})_{1/6}$, the weak doublet of scalar LQ's with hypercharge $Y=1/6$, which couples to SM fermions through a single gauge invariant operator, namely,~\cite{Becirevic:2015asa} 
\begin{align}
\label{eq:yuk-R2tilde}
\begin{split}
\mathcal{L}_{\widetilde{R}_2} &= -y_L^{ij}\, \overline{d_{Ri}} \widetilde{R}_2 i \tau_2 L_j+\mathrm{h.c.}\,,\\[0.4em]
&= -y_L^{ij}\, \overline{d_{Ri}} \ell_{Lj} \,\widetilde{R}_2^{(2/3)} + y_L^{ij}\, \overline{d_{Ri}} \nu_{Lj} \,\widetilde{R}_2^{(-1/3)}+\mathrm{h.c.}\,,
\end{split}
\end{align}

\noindent where $y_L$ is a generic matrix of Yukawa couplings. Another appealing feature of this scenario is that, like the $R_2$ model, the potentially troublesome diquark couplings to LQ are absent. Proton decay can still be generated in this scenario but by higher order operators which can be eliminated by imposing a suitable symmetry in a way similar to what has been done, for example, in Ref.~\cite{Cox:2016epl}. As before, we again identify the Wilson coefficients arising from the tree level contributions to $b\to s \ell_l^- \ell_k^+$ in this model and find,
\begin{equation}
 C_9^{\,\prime \, kl} =- C_{10}^{\,\prime \, kl} = - \dfrac{\pi v^2}{2 V_{tb}V_{ts}^\ast\alpha_{\mathrm{em}}}\dfrac{y_L^{sk} \big{(}y_L^{bl}\big{)}^\ast}{m_{\widetilde{R}_2}^2}\,,
\end{equation}

\noindent which turned out to be in agreement with $R_K^{\mathrm{exp}}<R_K^{\mathrm{SM}}$
but in conflict with $R_{K^{\ast}}^{\mathrm{exp}}<R_{K^{\ast}}^{\mathrm{SM}}$, cf.~discussion in Ref.~\cite{Becirevic:2015asa}. Furthermore, the Yukawa interactions in Eq.~\eqref{eq:yuk-R2tilde} do not contribute to the charged current processes, such as the transition $b\to c\tau \bar{\nu}$. This limitation can be overcome by introducing light right-handed neutrinos to this set-up~\cite{Becirevic:2016yqi}. In that case, scalar and tensor operators will be generated through the gauge invariant operator $\overline{Q}\widetilde{R}_2\nu_R$, which will not interfere with the SM contributions and therefore can provide only a small shift with respect to the SM predictions. 

		\subsubsection{$S_1=(\mathbf{\bar{3}},\mathbf{1})_{1/3}$}
	\label{sec:lq-list-S1}

Finally, one can also consider a scenario with $S_1= (\mathbf{\bar{3}},\mathbf{1})_{1/3}$, a weak singlet scalar LQ with hypercharge $Y=1/3$. This model was deemed to be viable in Ref.~\cite{Sakaki:2013bfa} for accommodating $R_{D^{(\ast )}}^{\mathrm{exp}}>R_{D^{(\ast )}}^{\mathrm{SM}}$. The most general Yukawa Lagrangian of $S_1$ reads
\begin{align}
\begin{split}
\mathcal{L}_{S_1} &= y_L^{ij} \, \overline{Q^C} i\tau_2 L_j\, S_1 + y_R^{ij} \,\overline{u^C_{R\,i}} e_{R\,j}\, S_1 +\mathrm{h.c.} \\[0.4em]
&= S_1 \Big{[}\big{(}V^\ast y_L \big{)}_{ij}\, \overline{u^C_{L\,i}}\ell_{L\,j}-y_L^{ij}\,\overline{d^C_{L\,i}}\nu_{L\,j}+y_R^{ij}\, \overline{u^C_{R\,i}}\ell_{R\,j} \Big{]} + \mathrm{h.c.}\,,
\end{split}
\end{align}

\noindent where $y_L$ and $y_R$ are generic Yukawa matrices. Like in the case of $S_3$ we omitted the terms involving diquark couplings to LQ, which must be forbidden 
by a symmetry to protect the proton stability. By integrating out the LQ state we obtain that, at the matching scale $\mu=m_{S_1}$,
\begin{align}
g_{V_L} &= \dfrac{v^2}{4 V_{cb}}\dfrac{y_L^{b\ell^\prime}\big{(}V y_L^\ast\big{)}_{c\ell}}{ m_{S_1}^2} \,, \\
g_{S_L} &= - 4 \, g_T = -\dfrac{v^2}{4 V_{cb}}\dfrac{y_L^{b{\ell^\prime}} \big{(}y_R^{c\ell}\big{)}^\ast}{ m_{S_1}^2}\,.
\end{align}

\noindent Although this LQ does not contribute to the transition $b\to s \ell_l^- \ell_k^+$ at tree-level, the effective coefficients $C_{9{(10)}}$ receive contributions at loop-level, namely~\cite{Bauer:2015knc}
\begin{align}
\label{eq:C9-S1}
\begin{split}
 C_{9}^{kl} +  C_{10}^{kl} &= \dfrac{m_t^2}{8 \pi \alpha_{\mathrm{em}}m_{S_1}^2}\big{(}V^\ast y_L\big{)}_{t k}\big{(}V^\ast y_L\big{)}_{t l}^\ast - \dfrac{v^2}{32\pi \alpha_{\mathrm{em}}m_{S_1}^2}\dfrac{\big{(}y_L\cdot y_L^\dagger\big{)}_{bs}}{V_{tb} V_{ts}^\ast}\big{(}y_L^\dagger\cdot y_L\big{)}_{kl}\,,\\[0.4em]
 C_{9}^{kl} - C_{10}^{kl} &= \dfrac{m_t^2}{8 \pi \alpha_{\mathrm{em}}m_{S_1}^2}\big{(}y_R\big{)}_{t k}\big{(}y_R\big{)}_{t l}^\ast\bigg{[}\log \dfrac{m_{S_1}^2}{m_t^2}-f(x_t)\bigg{]}\\
&\hspace*{4.7cm}- \dfrac{v^2}{32\pi \alpha_{\mathrm{em}}m_{S_1}^2}\dfrac{\big{(}y_L\cdot y_L^\dagger\big{)}_{bs}}{V_{tb} V_{ts}^\ast}\big{(}y_L^\dagger\cdot y_L\big{)}_{kl}\,,
\end{split}
\end{align}

\noindent where $x_t=m_t^2/m_W^2$ and $f(x_t)=1+\frac{3}{x_t-1}\left(\frac{\log x_t}{x_t-1}-1\right)$. The possibility of explaining $R_{K^{(\ast)}}$ and/or $R_{D^{(\ast)}}$ in this scenario will be discussed in Sec.~\ref{sec:which-LQ}.

	\subsection{Vector leptoquarks}
		\subsubsection{$U_1=(\mathbf{3},\mathbf{1})_{2/3}$}
		\label{sec:lq-list-U1}

The first scenario of this sort we consider is $U_1=(\mathbf{\bar{3}},\mathbf{1})_{2/3}$, the weak singlet vector LQ, which received considerable attention because it can provide a simultaneous explanation to the anomalies in $b\to s$ and $b\to c$ transitions~\cite{Buttazzo:2017ixm}. The most general Lagrangian consistent with the SM gauge symmetry allows couplings to both left-handed and right-handed fermions, namely,

\begin{equation}
\label{eq:lag-U1}
\mathcal{L}_{U_1} = x_L^{ij} \, \bar{Q}_i \gamma_\mu U_1^\mu L_j + x_R^{ij} \, \bar{d}_{R\,i} \gamma_\mu  U_1^\mu \ell_{R\,j}+\mathrm{h.c.},
\end{equation}

\noindent where $x_L^{ij}$ and $x_R^{ij}$ are the couplings. The contribution of the left-handed couplings to the effective Lagrangian~\eqref{eq:hamiltonian-bsll} amounts to

\begin{equation}
\label{eq:C9-U1}
 C_{9}^{kl} = - C_{10}^{kl} = -\dfrac{\pi v^2}{ V_{tb}V_{ts}^\ast \alpha_{\mathrm{em}}} \dfrac{x_L^{s l} \left(x_L^{b k}\right)^\ast}{m_{U_1}^2}\,,
\end{equation}

\noindent as required by the observation of $R_{K^{(\ast)}}^{\mathrm{exp}}<R_{K^{(\ast)}}^{\mathrm{SM}}$. Switching on the right-handed couplings, $x_R \neq 0$, amounts to contributions to other Wilson coefficients, $C_9^\prime = C_{10}^\prime$, $C_S = - C_P$ and $(C_S)^\prime = (C_P)^\prime$.  However, since the latter Wilson coefficients are disfavored by the current $b\to s$ data, we will set the right-handed couplings to zero and call such a scenario the minimal $U_1$ model. Furthermore, this scenario also contributes to $b\to c\ell \bar{\nu}_{\ell^\prime}$ by giving rise to the effective coefficient
\begin{align}
\label{eq:gV-U1}
\begin{split}
g_{V_L} &= \dfrac{v^2\,\left(V x_L\right)_{c\ell^\prime}\left(x_L^{b\ell}\right)^\ast}{ 2 V_{cb}\, m_{U_1}^2 }\\[0.3em]
&=  \dfrac{v^2}{2 m_{U_1}^2} \big{(}x_{L}^{b\ell}\big{)}^\ast\Big{[}x_L^{b\ell^\prime} + \dfrac{V_{cs}}{V_{cb}} x_L^{s\ell^\prime}+ \dfrac{V_{cd}}{V_{cb}} x_L^{d\ell^\prime}\Big{]}\,,
\end{split}
\end{align}

\noindent which is clearly acceptable since  the leading term implies $g_{V_L}>0$, in agreement with the observed 
$R_{D^{(\ast)}}^{\mathrm{exp}}>R_{D^{(\ast)}}^{\mathrm{SM}}$.

A peculiarity of this scenario is the absence of contributions to the transition $b\to s\nu\bar{\nu}$~\cite{Becirevic:2016oho,Buttazzo:2017ixm}. Importantly, however, this model is nonrenormalizable, which undermines its predictivity at the loop-level unless the ultraviolet (UV) mechanism generating the $U_1$ mass is explicitly specified, see Refs.~\cite{Assad:2017iib,DiLuzio:2017vat} for concrete examples.

		\subsubsection{$U_3=(\mathbf{3},\mathbf{3})_{2/3}$}
		\label{sec:lq-list-U3}

The last scenario we consider is the weak triplet of vector LQ's, $U_3=(\mathbf{3},\mathbf{3})_{2/3}$~\cite{Fajfer:2015ycq}. Due to gauge symmetry, this LQ couples only to the left-handed SM fermion doublets, with the most general Lagrangian being~\cite{Dorsner:2016wpm}

\begin{equation}
\label{eq:lag-U3}
\mathcal{L}_{U_3} = x_L^{ij} \, \overline{Q}_i \gamma^\mu (\tau_k U_{3\,\mu}^k) L_j +\mathrm{h.c.}
\end{equation}

\noindent Using Eq.~\eqref{eq:lag-U3} and matching at tree-level onto the effective Lagrangian in Eq.~\eqref{eq:hamiltonian-bsll}, we obtain that

\begin{equation}
\label{eq:C9-U3}
 C_{9}^{kl} = -  C_{10}^{kl} = -\dfrac{\pi v^2}{ V_{tb}V_{ts}^\ast \alpha_{\mathrm{em}}} \dfrac{x_L^{s l} \left(x_L^{b k}\right)^\ast}{m_{U_3}^2}\,,
\end{equation}

\noindent which can accommodate $R_{K^{(\ast)}}$, in a way similar to the $U_1$ model. Concerning the effective operators in the charged-current case~\eqref{eq:lagrangian-lep-semilep}, the only non-vanishing effective coefficient is again $g_{V_L}$ which is given by

\begin{align}
\label{eq:gV-U3}
\begin{split}
g_{V_L} &= - \dfrac{v^2\,\left(V x_L\right)_{c\ell^\prime}\left(x_L^{b\ell}\right)^\ast}{ 2 V_{cb}\, m_{U_3}^2 }\,,\\
&=-\dfrac{v^2}{2 m_{U_3}^2}\left(x_L^{b\ell}\right)^\ast \Big{[}x_L^{b\ell^\prime} + \dfrac{V_{cs}}{V_{cb}} x_L^{s\ell^\prime}+ \dfrac{V_{cd}}{V_{cb}} x_L^{d\ell^\prime}\Big{]}\,.
\end{split}
\end{align}

\noindent We see that, similar to the scenario with $S_3$, this model cannot accommodate the deviation in $R_{D^{(\ast)}}$ because the term proportional to $|x_L^{b\tau}|^2$ is negative, while the others are tightly constrained by other flavor physics observables, such as $\mathcal{B}(B\to K \nu\bar{\nu})$. 

Similarly to $U_1$, the $U_3$ model is generally nonrenormalizable. Nevertheless, under certain circumstances, loops involving $U_3$ can be calculated. More precisely, if the $3 \times 3$ matrix $x_L$ from Eq.~\eqref{eq:lag-U3} is unitary, UV-divergences appearing in loop-induced FCNCs mediated by $U_3$ are canceled through the GIM mechanism. However, the price to pay for having a unitary coupling matrix is that LQ couplings to first generation SM fermions, such as $e$, $d$, or $u$, can no longer be avoided. In turn, the presence of such couplings is in strong conflict with LFV bounds from $\mu - e$ conversion in Au nuclei and from $\mathcal{B}(K_L \to \mu e)$, which exclude the $U_3$ scenario with unitary $x_L$ as a viable explanation of the $b \to s$ anomalies, see discussion in Ref.~\cite{Becirevic:2016oho}.

\section{High-$p_T$ phenomenology}
	\label{sec:LHC}

	\subsection{Direct limits on pair-produced LQs}
	\label{sec:lhc- pair-lq}

\begin{figure}[b!]
  \centering
  \includegraphics[width=0.75\textwidth]{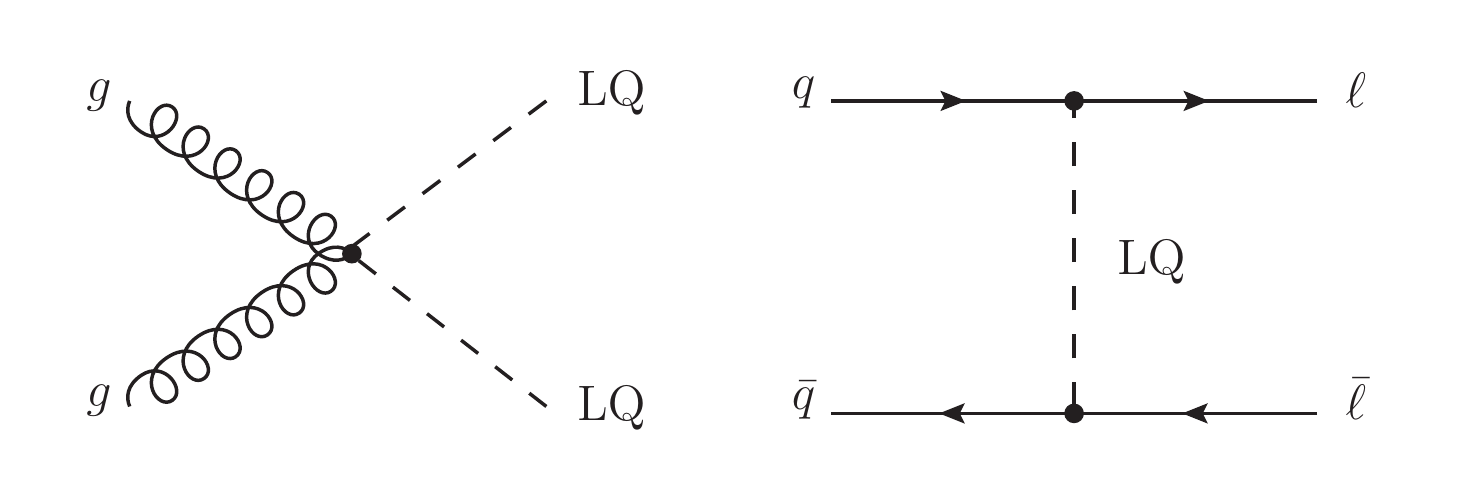}\\
  (a)\hspace{5.5cm}(b)
  \caption{ \sl \small (a) Representative Feynman diagram for LQ pair production via QCD interactions. (b) Feynman diagram for LQ $t$-channel exchange in $pp\to\ell\bar\ell$ production at the LHC. The dashed propagator represents either a scalar or vector LQ state.}
  \label{fig:LQ-diagrams}
\end{figure}

An efficient way to set limits on LQs is to directly search for them at hadron colliders. At the LHC one of the most significant example of such a processes is the pair production $gg\,(q\bar q)\to \text{LQ}^\dagger\text{LQ}$, shown in Fig.~\ref{fig:LQ-diagrams}\,(a). In both ATLAS and CMS the searches for this process in different decay channels into second and/or third generation quarks and leptons, $\mathrm{LQ}^\dagger\mathrm{LQ}\to q\bar q \ell\bar \ell,\,q\bar q\nu\bar \nu$, have been made. The results of these searches lead to model independent bounds on both the mass and branching fractions of the LQ.

\begin{table}[t]
\renewcommand{\arraystretch}{1.5}
\centering
\begin{tabular}{|c|c|c|c|c|c|}
\hline 
Decays& LQs  & Scalar LQ limits & Vector LQ limits&  $\lumi$ / Ref. \\\hline\hline
 $ j j\,\tau\bar\tau$     & $S_1,R_2,S_3,U_1,U_3$ & -- & -- & -- \\ 
 $ b\bar b\,\tau\bar\tau$ & $R_2,S_3,U_1,U_3$     & 850 (550)~GeV & 1550 (1290)~GeV & 12.9~\invfb\cite{Sirunyan:2017yrk} \\ 
 $ t\bar t\,\tau\bar\tau$ & $S_1,R_2,S_3,U_3$     & 900 (560)~GeV & 1440 (1220)~GeV & 35.9~\invfb\cite{Sirunyan:2018nkj} \\ \hline
 $ j j\,\mu\bar\mu$           & $S_1,R_2,S_3,U_1,U_3$ & 1530 (1275)~GeV & 2110 (1860)~GeV & 35.9~\invfb\cite{CMS:2018sgp}\\ 
 $ b\bar b\,\mu\bar\mu$   & $R_2,U_1,U_3$          & 1400 (1160)~GeV & 1900 (1700)~GeV & 36.1~\invfb\cite{Diaz:2017lit} \\  
 $ t\bar t\,\mu\bar\mu$   &    $S_1,R_2,S_3,U_3$  &  1420 (950)~GeV &  1780  (1560)~GeV & 36.1~\invfb\cite{Camargo-Molina:2018cwu,CMS:2018itt}\\  \hline
 $ j j\,\nu\bar\nu$           & $R_2,S_3,U_1,U_3$     & 980 (640)~GeV  & 1790 (1500)~GeV & 35.9~\invfb\cite{CMS:2018bhq} \\ 
 $ b\bar b\,\nu\bar\nu$   & $S_1,R_2,S_3,U_3$     & 1100 (800)~GeV & 1810 (1540)~GeV & 35.9~\invfb \cite{CMS:2018bhq} \\ 
 $ t\bar t\,\nu\bar\nu$    & $R_2,S_3,U_1,U_3$     & 1020 (820)~GeV & 1780 (1530)~GeV & 35.9~\invfb \cite{CMS:2018bhq} \\ 
    \hline
\end{tabular}
\caption{ \sl \small Summary of the current limits from LQ pair production searches at the LHC. In the first column we give the searched final states and in the second column the LQs for which this search is relevant. In the next two columns we present the current limits on the mass for scalar and vector LQs, respectively, for $\beta=1~(\beta=0.5)$. In the last column we display the value of the LHC luminosity for each search along with the experimental references. Note that ``$j$" denotes any jet originating from a charm or a strange quark. Entries marked with ``$-$" indicate that no recast or search in this channel has been performed up to this date.}
\label{tab:LQ-pair-bounds} 
\end{table}

 In Table~\ref{tab:LQ-pair-bounds} we list the most recent lower limits on the masses of second/third generation scalar and vector LQs relevant to this work, for benchmark branching ratios set to $\beta\!=\!1\,(0.5)$. These limits assume the following: (i) pair production is dominated by QCD interactions, and (ii) for vector LQs ($V^\mu$) the LQ-gluon interaction term, $\mathcal{L}\supset -\kappa g_s V^\mu G_{\mu\nu} V^\nu$, is taken with $\kappa=1$. The first assumption is in general true for LQ-fermion couplings of order $\sim1$ or smaller~\cite{Dorsner:2014axa}. In this regime, contributions to $q\bar q\to\mathrm{LQ}^\dagger\mathrm{LQ}$ with a $t$-channel lepton (where the amplitude is proportional to the squared LQ-fermion coupling) are subleading compared to QCD induced production. The assumption on the value of $\kappa$, instead, depends on the UV origin of the vector LQ~\cite{Blumlein:1996qp}. If $V^\mu$ is a fundamental gauge boson of a new non-abelian gauge group then the gauge symmetry completely fixes the choice $\kappa=1$. The possibility of having $|\kappa|<1$ may arise in a UV theory where the vector LQ is a composite particle, therefore giving rise to LHC limits weaker than for the gauge vector LQ presented in Table.~\ref{tab:LQ-pair-bounds}.

	\subsection{Limits from high-$p_T$ tails of $pp\to \ell\bar\ell$}
	\label{sec:lhc-dilepton}


As shown in Refs.~\cite{Faroughy:2016osc,Greljo:2017vvb}, a contribution arising from the $t$-channel exchange of LQs to $pp\to\ell\bar\ell$ ($\ell=\mu,\tau$) can be directly probed in the high-$p_T$ tails of Drell-Yan processes at the LHC. In particular, larger values of Yukawa couplings, that are often needed to accommodate the $B$-anomalies, could modify the tail of the differential cross section of $pp\to \ell\bar\ell$. In the following we use LHC data from $pp$-collisions at 13~TeV to set limits on each LQ model. For this we have recast two recent searches by ATLAS at 36.1~fb$^{-1}$ for a $Z^\prime$ decaying to $\mu\bar\mu$ \cite{Aaboud:2017buh} and $\tau\bar\tau$ \cite{Aaboud:2017sjh}, respectively. 

 For the di-tau analysis we focused on the fully inclusive channel with hadronic taus ($\hadtau$) in the final state, given that these perform considerably better at high $p_T$ than the leptonic tau decay channels. For each search we counted the number of observed and background events above different threshold values of the invariant mass distributions $m_{\mu\mu}$ for the di-muon search and the total transverse mass distribution $m_{\text{tot}}$ for the di-tau search (see ref.~\cite{Aaboud:2017sjh} for the definition of $m_{\text{tot}}$). An upper bound at 95\% C.L. on the number of allowed signal events above each mass threshold was extracted for each search by minimizing the Log-Likelihood ratio with nuisance parameters for the background uncertainties, as described in \cite{Cowan:2010js}. Besides the current luminosity limits, we also estimated projected limits at a higher LHC luminosity of $\lumi\!=\!300$~\invfb by scaling the data and background events with the luminosity ratio and the background uncertainties with $\sqrt{\lumi}$ assuming that the data in the distribution tails are statistically dominated. This assumption holds well for the leading backgrounds such as SM Drell-Yan production or fake $\hadtau$ from QCD jet mis-tagging since these are estimated using experimental data from control regions that improve with more statistics. Additionally, systematics in the tails of the di-muon and di-tau searches are well under control and only dominate the lower bins where the search is insensitive to the massive LQs. 
 

%
\begin{figure}[t!]
  \centering
  \includegraphics[width=0.5\textwidth]{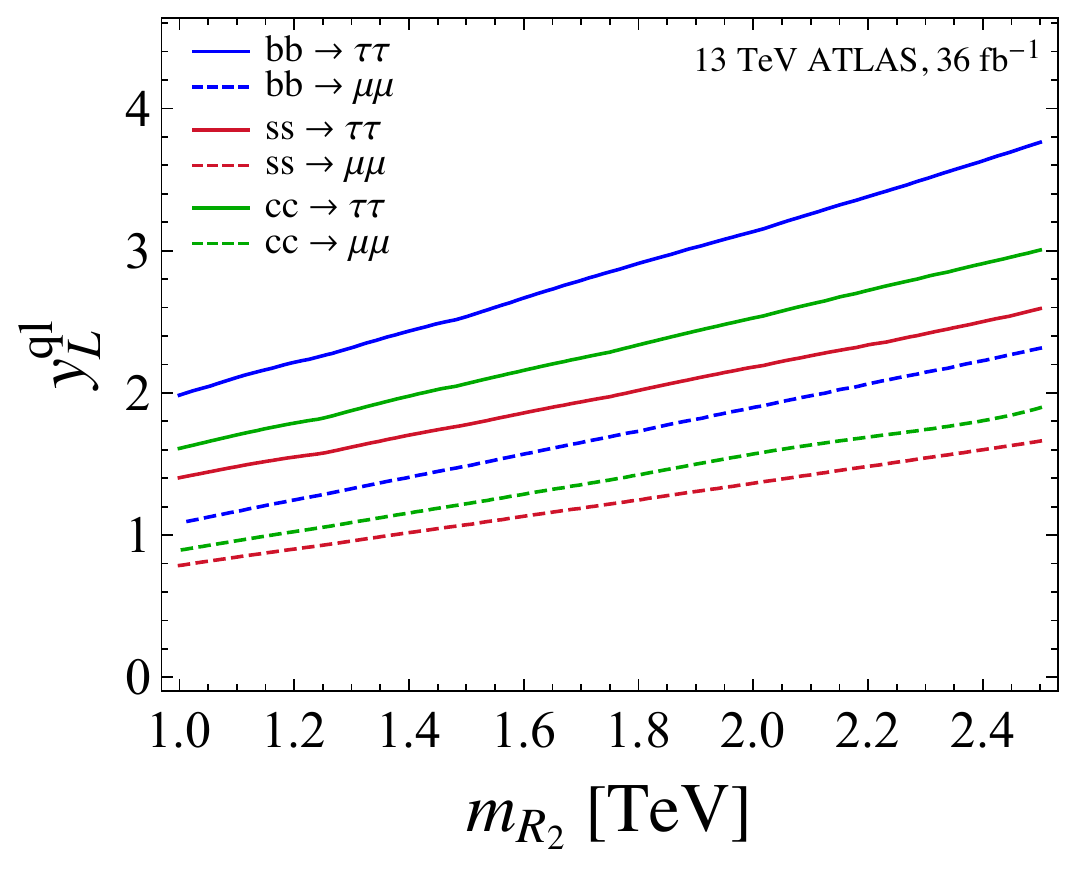}~  \includegraphics[width=0.5\textwidth]{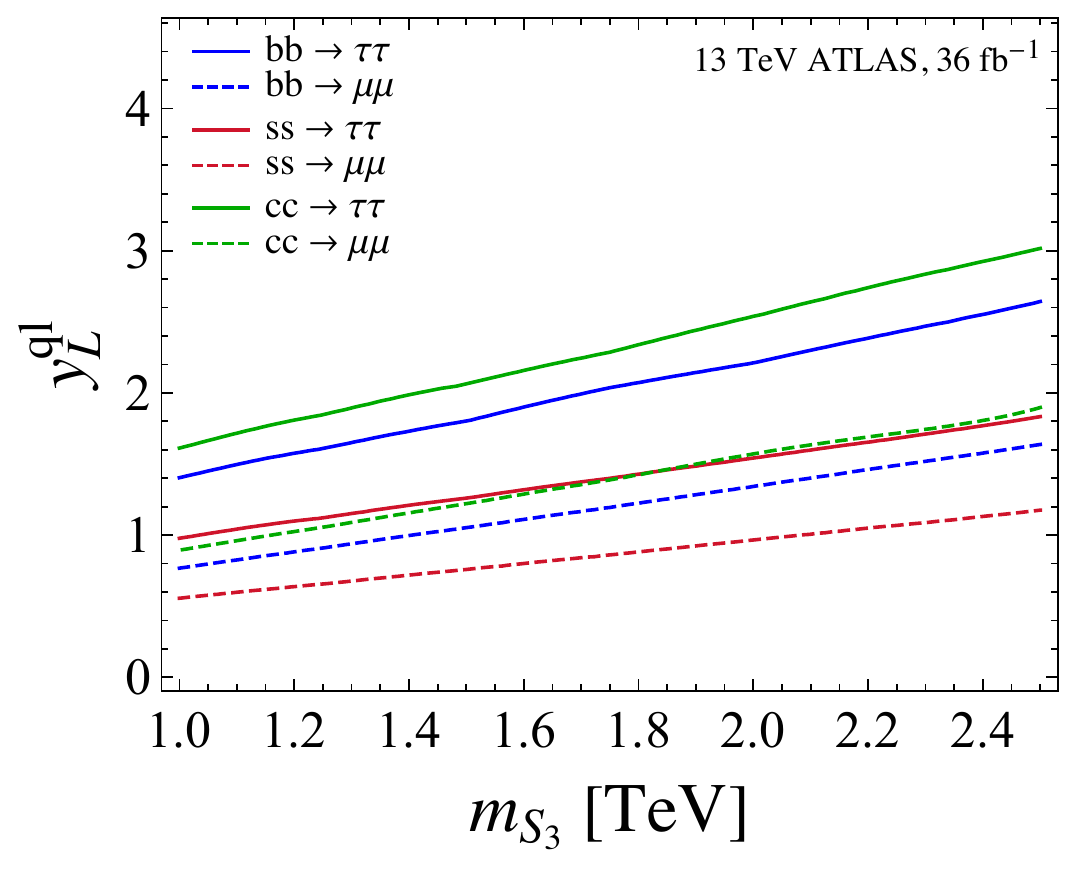}
  \includegraphics[width=0.5\textwidth]{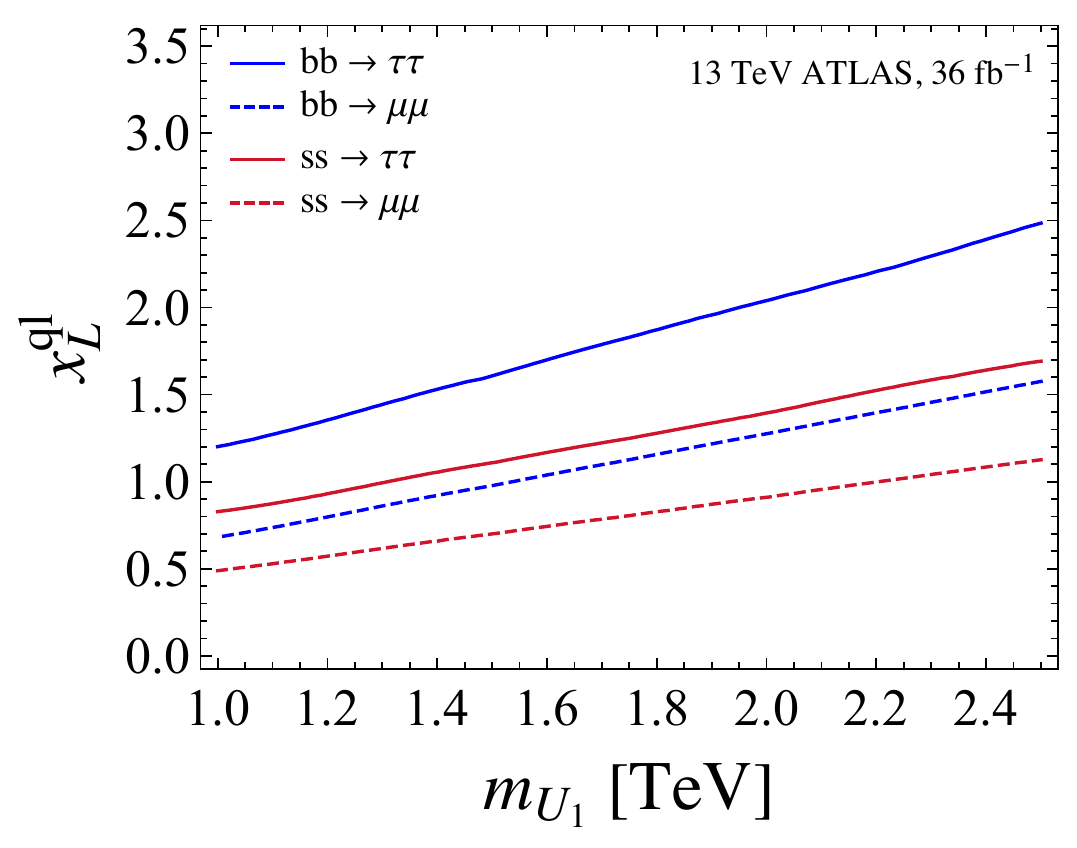}~  \includegraphics[width=0.5\textwidth]{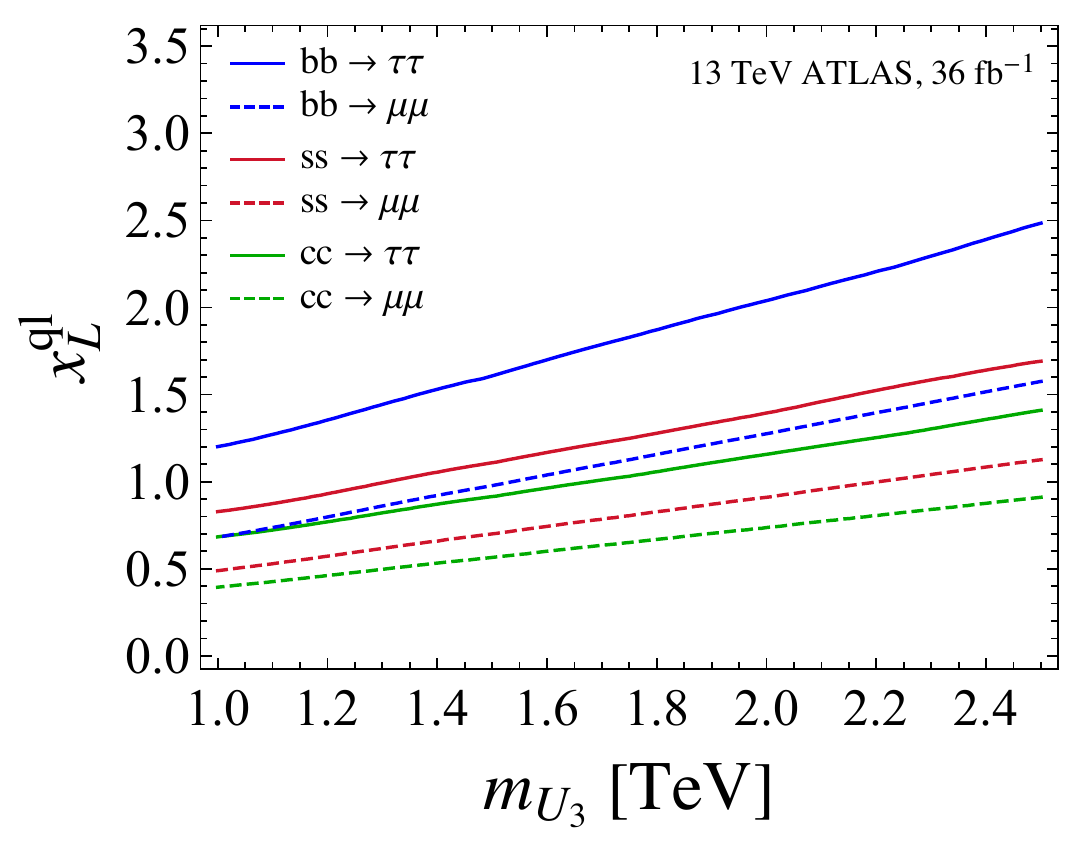}
  \caption{ \sl \small The top panel (lower panel) shows current limits in the coupling vs mass plane for several scalar LQ (vector LQ) models from LHC searches in $pp\to\ell\bar\ell$ high-$p_T$ tails at 13~TeV with $36$\,\invfb of data. The solid  and dashed lines represent limits from di-tau and di-muon searches, respectively, for different initial quarks while turning one scalar (vector) LQ coupling  $y^{ql}_L$ ($x^{ql}_L$) at a time.}
  \label{fig:fdileptonLHC}
\end{figure}

 For our simulations we created the Universal FeynRules Output (UFO) files using {\tt FeynRules}~\cite{Alloul:2013bka} for each LQ mediator ($S_1$, $S_3$, $R_2$, $U_1$ and $U_3$) coupling exclusively to second and third generation of quarks and leptons. For $R_2$ and $U_1$ we have only considered, for simplicity, non-zero left-handed Yukawa matrices in Eq.~\eqref{eq:yuk-R2-bis} and Eq.~\eqref{eq:lag-U1}. After exporting the UFOs to {\tt MadGraph5} \cite{Alwall:2014hca} we generated for each LQ mediator a statistically significant set of $t$-channel Drell-Yan event samples $q\bar q\to\mu\bar\mu,\,\tau\bar\tau$ for initial sea quarks, $q\in \{s,c,b\}$, and for vector (scalar) LQs at different masses in the 1-6 TeV (0.6-3 TeV) range. Each sample was subsequently showered and hadronized using {\tt Pythia8}~\cite{Sjostrand:2014zea}. Final state hadronic taus and isolated muons where reconstructed and smeared using {\tt Delphes3}~\cite{deFavereau:2013fsa} with the parameters set according to each experimental scenario. 
 In order to illustrate the current reach of the LHC for each LQ state we show in Fig.~\ref{fig:fdileptonLHC} results from the $pp\to\mu\bar\mu$ and $pp\to\tau\bar\tau$ searches by ATLAS for scalar (vector) LQs in the $y_L^{ql}$ ($x_L^{ql}$) coupling versus mass plane.~\footnote{We did not present plots for $S_1$ since these bounds are identical to the bounds for $S^{1/3}_3$.}  Each line corresponds to the $95\%$ upper limit for the process $q\bar q\to\ell\bar\ell$ turning on each flavor coupling one at a time, with $q\in\{s,c,b\}$ and $\ell\in\{\mu,\tau\}$. Similar bounds of the same order can be extracted for the coupling products $y^{ql}_L y^{q^\prime l}_L$ and $x^{ql}_Lx^{q^\prime l}_L$ with $q\ne q^\prime$ from the quark flavor violating process $qq^\prime\to\ell\bar\ell$. 
 
Besides producing deviations in the di-lepton tails, LQ mediators that couple simultaneously to differently charged leptons may also produce measurable effects in LFV observables at the LHC. In particular, searching for the process $pp\to \ell\ell^\prime$ with a LQ exchanged in the $t$-channel may provide an additional handle for setting constraints on the flavor structure of these LQ models. Existing searches for a massive $Z^\prime$ with LFV couplings have been presented by the LHC collaborations in the $Z^\prime\to e\mu,\, e\tau,\,\mu\tau$ channels. In order to determine the sensitivity of the LHC to the $t$-channel process $pp\to \mu\tau$ we recast the LFV $Z^\prime$ search by ATLAS \cite{Aaboud:2016hmk} at $3.2$\,\invfb. We find, however,  that the bounds on the LQs extracted from the high-$p_T$ $m_{\mu\tau}$ tails are always weaker than the combined bounds from the flavor conserving di-muon and di-tau tails described above.

\section{Which leptoquark model?}
	\label{sec:which-LQ}
	
In this Section we subject the models listed in Sec.~\ref{sec:lq-list} to the constraints stemming both from direct and indirect NP searches. We will then select LQ scenarios which can accommodate $R_{K^{(\ast)}}^\mathrm{exp}$ and/or $R_{D^{(\ast)}}^\mathrm{exp}$  based on the expressions derived above. 

\begin{itemize}
	\item[•]$S_3=(\mathbf{\bar{3}},\mathbf{3})_{1/3}$\\[0.4em] 
	As already discussed in Sec.~\ref{sec:lq-list-S3}, the $S_3$ model can accommodate $R_{K^{(\ast)}}^\mathrm{exp}$ since it predicts the NP contribution to $C_{9,10}$ satisfying $C_9=- C_{10}$. On the other hand, this scenario cannot accommodate the anomalies in $R_{D^{(\ast)}}^\mathrm{exp} > R_{D^{(\ast)}}^\mathrm{SM}$. This can be easily understood from Eq.~\eqref{eq:gV-S3}, where we see that the sign of the leading term for $g_{V_L}$ is negative, thus further lowering the value of $R_{D^{(\ast)}}^{\mathrm{SM}}$, see also~Eq.~\eqref{eq:RDfit}. The subleading terms in this equation could in principle provide a positive contribution to $R_{D^{(\ast)}}$ but such a situation would be in conflict with tight constraints coming from other flavor physics limits, such as $B\to K \nu\nu$ and $\Delta m_{B_s}$. 
	
	\item[•]$R_2=(\mathbf{{3}},\mathbf{2})_{7/6}$:\\[0.4em]
	 At tree level the $R_2$ model contributes to $C_{9,10}$ in such a way that they satisfy $ C_9 = C_{10}$, which disagrees with $R_{K^{(\ast)}}< R_{K^{(\ast)}}^{\mathrm{SM}}$, cf.~Sec.~\ref{sec:lq-list-R2}. This situation can be avoided by choosing a flavor structure such that the tree-level contribution to $b\to s\mu\mu$ is absent. As a result,  $C_{9,10}$ are loop induced and the NP Wilson coefficients satisfy $C_9 =-C_{10}$, as needed~\cite{Becirevic:2018uab}. The minimal flavor ansatz in Eq.~\eqref{eq:yuk-R2-bis} to realize such a scenario is given by

\begin{equation}
\label{eq:yL-R2-pattern}
y_L  = \begin{pmatrix}
0 & 0 & 0\\ 
0 & y_L^{c\mu} & 0 \\ 
0 & y_L^{t\mu} & 0
\end{pmatrix}\,, \qquad\qquad y_R = 0\,,
\end{equation}
	
where $y_{L}^{c\mu}$ and $y_{L}^{t\mu}$ are non-zero couplings. For the time being, we neglected the coupling to $\tau$ because it plays no role in the discussion of $R_{K^{(\ast)}}$.  The main challenge in this scenario is to comply with LHC limits, which are particularly constraining in the region the anomalies can be accounted for. To see the extent to which the model can accommodate measured $R_{K^{(\ast)}}$ we use the allowed values of Wilson coefficients specified in Eq.~\eqref{eq:C9-exp}, as well as the constraints arising from 
the limits on $\mathcal{B}({B\to K^{(\ast)}\nu \bar{\nu}})$ and from the $Z$-pole observables~\cite{Becirevic:2017jtw}. 
The results of our fit are then combined with limits derived in Sec.~\ref{sec:LHC},  and shown in Fig.~\eqref{fig:coupl-R2}, for two benchmark masses, $m_{R_2}=1.2$~TeV and $1.5$~TeV, cf.~Table~\ref{tab:LQ-pair-bounds}. From this plot we see that a large $y_L^{c\mu}$ needed at low-energies is in tension with LHC limits, allowing then an explanation of $R_{K^{(\ast)}}$ only at the $2\sigma$ level. Therefore, if the central values of $R_K^{\mathrm{exp}}$ and $R_{K^\ast}^{\mathrm{exp}}$ remain unchanged with more data, this model will be excluded as an explanation of the $b\to s$ anomalies. A similar conclusion has been independently found in Ref.~\cite{Camargo-Molina:2018cwu}.

\begin{figure}[tbp!]
  \centering
 \includegraphics[width=0.5\textwidth]{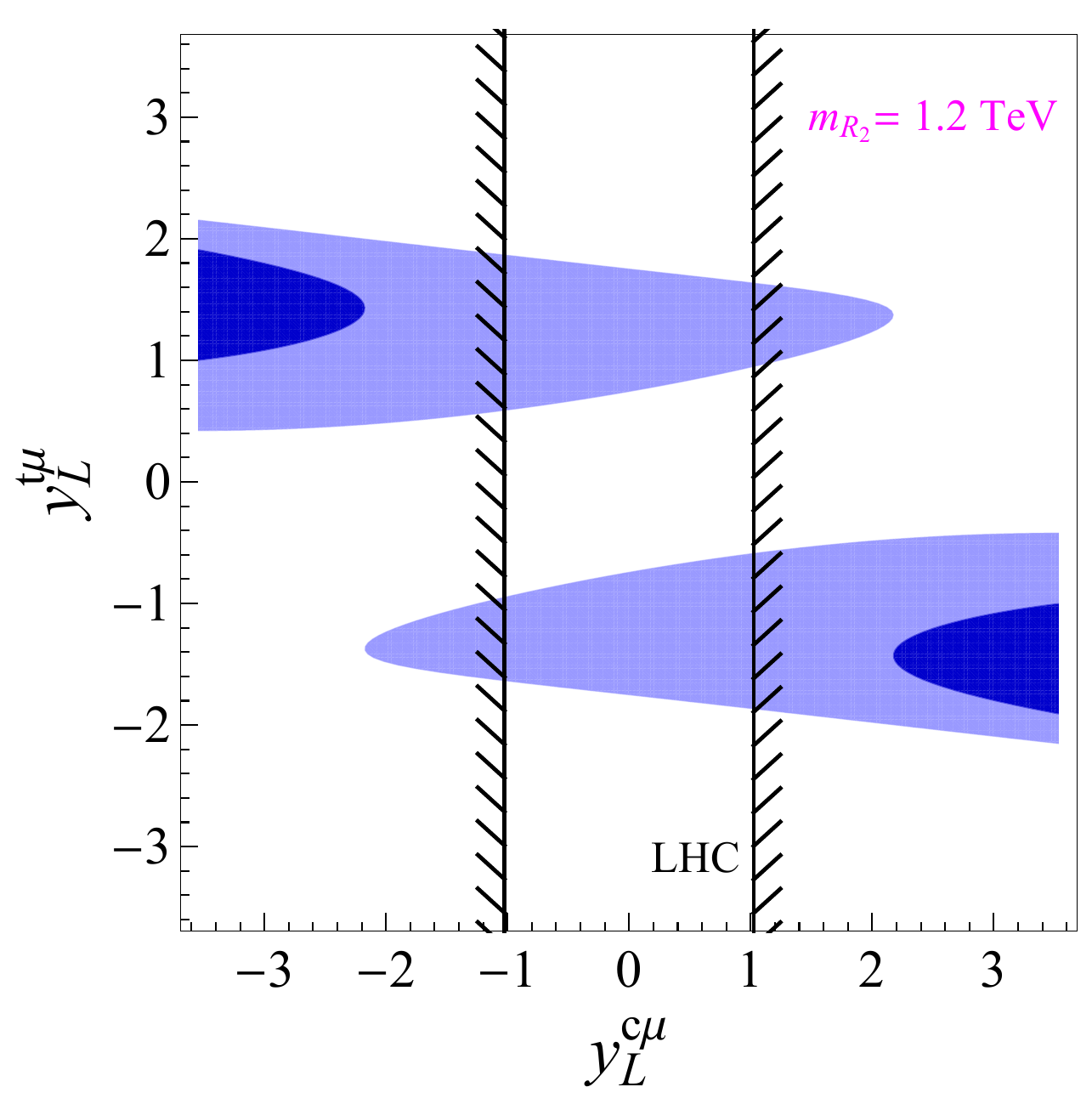}~\includegraphics[width=0.5\textwidth]{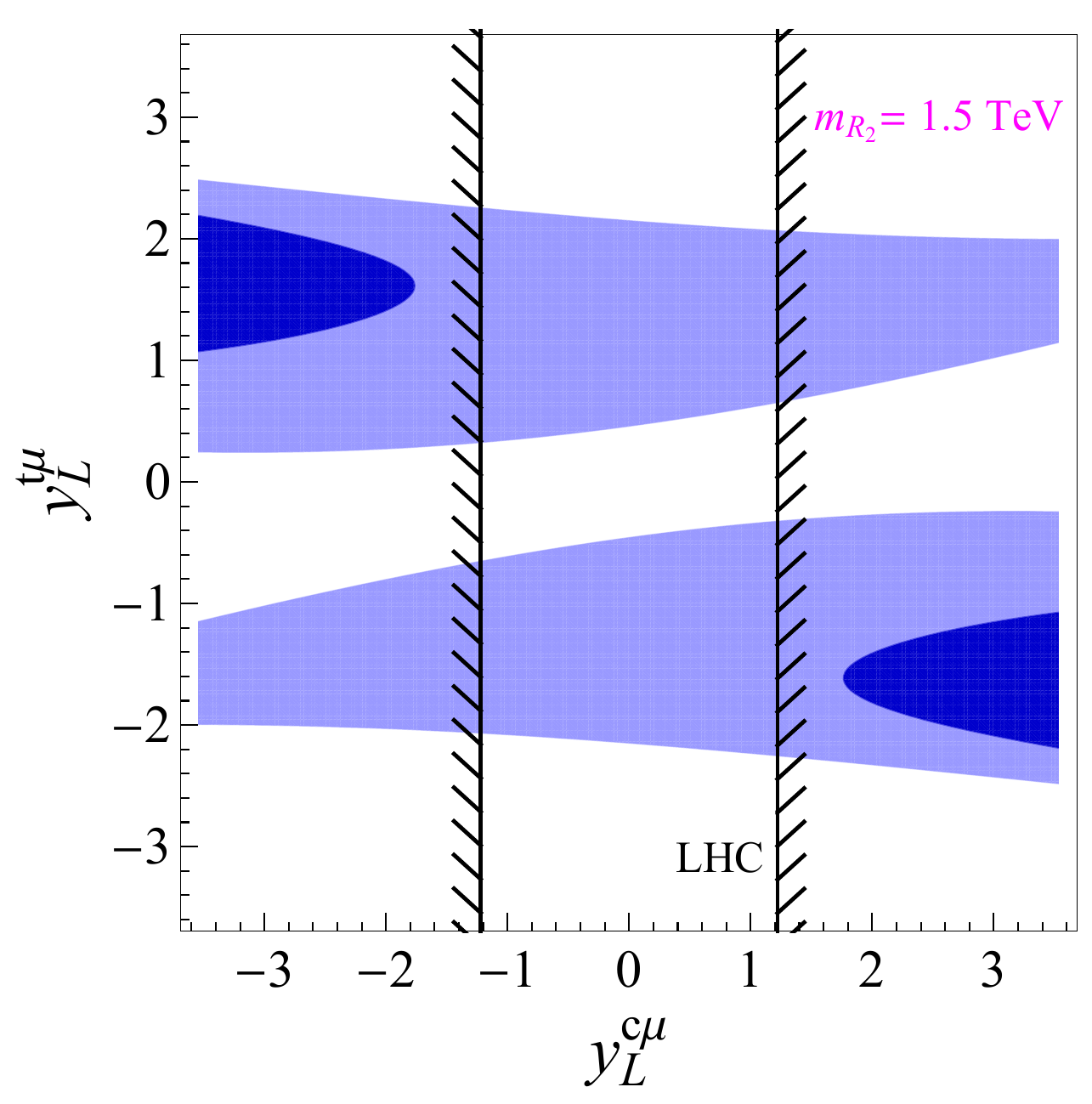}
  \caption{ \sl \small Regions in plane $y_{c\mu}$ vs.~$y_{t\mu}$ with $\Delta \chi^2 =2.3 (6.8)$ are plotted in dark (light) blue for $m_{R_2}=1.2$~TeV (left panel) and 1.5$~TeV$ (right panel). We consider the $b\to s\ell\ell$ constraint [Eq.~\eqref{eq:C9-exp}] and the ones from the $Z$-pole observables computed in~\cite{Becirevic:2017jtw}. The black line denotes the LHC limits derived from $pp\to \mu\mu$ data with $36~\mathrm{fb}^{-1}$, which excludes an explanation of $R_{K^{(\ast)}}$ to $1\sigma$ for both masses. See text for details.}
  \label{fig:coupl-R2}
\end{figure}

The above flavor pattern is clearly not satisfactory to explain the observed $R_{D^{(\ast)}}>R_{D^{(\ast)}}^\mathrm{SM}$. To accommodate those one could 
proceed as in Refs.~\cite{Becirevic:2018uab,Becirevic:2018afm,Sakaki:2013bfa,Hiller:2016kry} and let $y_{L}^{c\tau} \,\big{(}y_R^{b\tau}\big{)}^\ast$ to be of order $\mathcal{O}(1)$, which is compatible both with  low-energy observables and with upper limits on $pp\to\tau\tau$ derived at LHC. This is illustrated in Fig.~\ref{fig:fdileptonLHC}. From the point of view of the effective Lagrangian~\eqref{eq:lagrangian-lep-semilep} this LQ scenario generates the combination $g_{S_L} = 4\, g_T$ at the matching scale $\mu\simeq 1~\mathrm{TeV}$, which can turn $R_{D^{(\ast)}}$  compatible with $R_{D^{(\ast)}}^{\mathrm{exp}}$ if $g_{S_L}$ is mostly imaginary, cf.~Fig.~\ref{fig:fits-RD}. We should reiterate that a simultaneous explanation of both $R_{K^{(\ast)}}$ and $R_{D^{(\ast)}}$ cannot be obtained in this scenario even at the $2\sigma$ level. This is so because the couplings required to accommodate each of the observed anomalies would induce too large a value for $\mathcal{B}(\tau \to \mu \gamma)$, due to the chiral enhancement by the top quark~\cite{Becirevic:2017jtw}. 
	
		\item[•]$\widetilde{R}_2=(\mathbf{{3}},\mathbf{2})_{1/6}$\\[0.4em]
In a model with the LQ state $\widetilde{R}_2$, the non-zero NP Wilson coefficients in the $b\to s\mu\mu$ decay are those corresponding to the chirality flipped operators and they satisfy $C_9^\prime = -  C_{10}^\prime$. While this is enough to explain $R_K < R_K^\mathrm{SM}$, it predicts $R_{K^\ast}>R_{K^\ast}^{\mathrm{SM}}$ which is in disagreement with experimental data~\cite{Becirevic:2015asa}. In Ref.~\cite{Becirevic:2016yqi} this model has also been considered to account for the anomalies in charged currents. To this purpose, it is necessary to postulate the existence of light right-handed neutrinos, as already mentioned in Sec.~\ref{sec:lq-list-R2tilde}. The main difficulty of this scenario is to face the LHC limits on (i) recent limits on pair-produced LQs given in Table~\ref{tab:LQ-pair-bounds}, and (ii) $\tau\tau$ dilepton tails~\cite{Faroughy:2016osc}. The combination of these latter constraints is particularly efficient in this scenario, so that $\widetilde{R}_2$ model can only account for a very small enhancement in $R_{D^{(\ast)}}$.

		\item[•]$S_1=(\mathbf{{\bar{3}}},\mathbf{1})_{1/3}$\\[0.4em]
The $S_1$ model is a viable candidate to explain $R_{D^{(\ast)}}> R_{D^{(\ast)}}^\mathrm{SM}$ since it generates the effective couplings $g_{V_L}$ and $g_{S_L}=-4\, g_T$, both being equally viable when it comes down to accommodating $R_{D^{(\ast)}}^\mathrm{exp}> R_{D^{(\ast)}}^\mathrm{SM}$, see~Fig.~\ref{fig:fits-RD}. On the other hand, this scenario does not provide a tree-level contribution to $b\to s\ell\ell$ but $C_{9,10}$ can be induced by the box-diagrams satisfying $C_9=-C_{10}$, as desired~\cite{Bauer:2015knc}. 
It appears, however, that this is not a viable scenario if one works with $m_{S_1} \approx 1$~TeV and if one wants to stay compatible with $R_{D^{(\ast)}}^{\mu/e} = \mathcal{B}(B\to D^{(\ast)} \mu\bar{\nu})/\mathcal{B}(B\to D^{(\ast)} e \bar{\nu})$, as discussed in Ref.~\cite{Becirevic:2016oho}.  A step beyond was made in Ref.~\cite{Cai:2017wry} in which the model was shown to be viable if (i) a larger LQ mass is considered, and (ii) a different coupling structure, with large right-handed LQ couplings to the $\tau$-lepton, is allowed. 
	
To evaluate the extent to which this particle can improve the description of the $b\to s$ anomalies, we consider the minimal flavor ansatz,

\begin{equation}
\label{eq:xL-S1-pattern}
y_L  = \begin{pmatrix}
0 & 0 & 0\\ 
0 & y_L^{s\mu} & 0 \\ 
0 & y_L^{b\mu} & 0
\end{pmatrix}\,, \qquad\qquad y_R = 0\,,
\end{equation}

\noindent where we keep only the couplings that contribute to $C_9=-C_{10}$.~\footnote{Note that the couplings $y_R^{ij}$ to right-handed leptons will generate the combination of Wilson coefficients $C_9+C_{10}$, which is disfavored by the current data.} We then perform a fit to the low-energy observables modified by these couplings, namely, (i) the values of Wilson coefficients given in~Eq.~\eqref{eq:C9-exp}, (ii) the experimental limits on $\mathcal{B}(B\to K \nu \bar{\nu})$, (iii) the experimentally established $\Delta m_{B_s}$, (iv) $Z$-boson couplings to leptons measured at LEP, (v) leptonic decays $D_{(s)}\to \mu \bar{\nu}$ and $K\to \mu\bar{\nu}$, and (vi) the rare charm decay $D^0\to\mu\mu$. All the expression needed for that analysis, together with the experimental values/bounds can be found in Ref.~\cite{Becirevic:2016oho}. We find $\chi^2_{\mathrm{min}}/\mathrm{d.o.f.}=11.2/3$ for the following choice of parameters:

\begin{equation}\label{eq:S1-fit}
y_L^{b\mu}\approx \sqrt{4\pi}\,,\qquad\quad y_L^{s\mu}\approx -0.15\,,\qquad \quad m_{S_1} \approx 5.2~\mathrm{TeV}\,,
\end{equation}

\noindent where we have imposed the perturbativity limit $|y_L^{ij}| \leq \sqrt{4\pi}$. The value $\chi^2_{\mathrm{min}}=11.2$ is smaller than the SM value, $\chi^2_{\mathrm{SM}} = 21.2$, showing that the discrepancy can be decreased but not fully accommodated. In Fig.~\ref{fig:coupl-S1-muons} we plot the regions with $\Delta \chi^2 <2.3$ and $6.8$ in the planes $m_{S_1}$ vs.~$y_L^{s\mu}$ and $m_{S_1}$ vs.~$y_L^{b\mu}$. From this plot we see that one needs large LQ masses and large $|y_L^{b\mu}|$ to satisfactorily accommodate the current data. Such large couplings are close to the nonperturbative regime $|y_L^{b\mu}|\gtrsim \sqrt{4\pi}$, indicating that one needs a systematic study of one-loop corrections before assessing the viability of this scenario. In Fig.~\ref{fig:coupl-S1-muons}  we also show the bounds coming from the direct searches at LHC which are weak in the region of parameters described above. By assuming $|y_L^{b\mu}|\gg |y_L^{s\mu}|$, the bound we deduce from the pair production searches in the $t\bar t\mu\bar\mu$ channel is $m_{S_1}>1.4$~TeV \cite{CMS:2018itt}, much lower than the one we get from the flavor fit, given in Eq.~\eqref{eq:S1-fit}. We should note that limits from the study of dilepton tails, as described in Sec.~\ref{sec:lhc-dilepton}, are too weak, as expected, because the flavor fit favors couplings mostly to the third generation of quarks, for which the PDFs are very suppressed.

\begin{figure}[htbp!]
  \centering
  \includegraphics[width=0.5\textwidth]{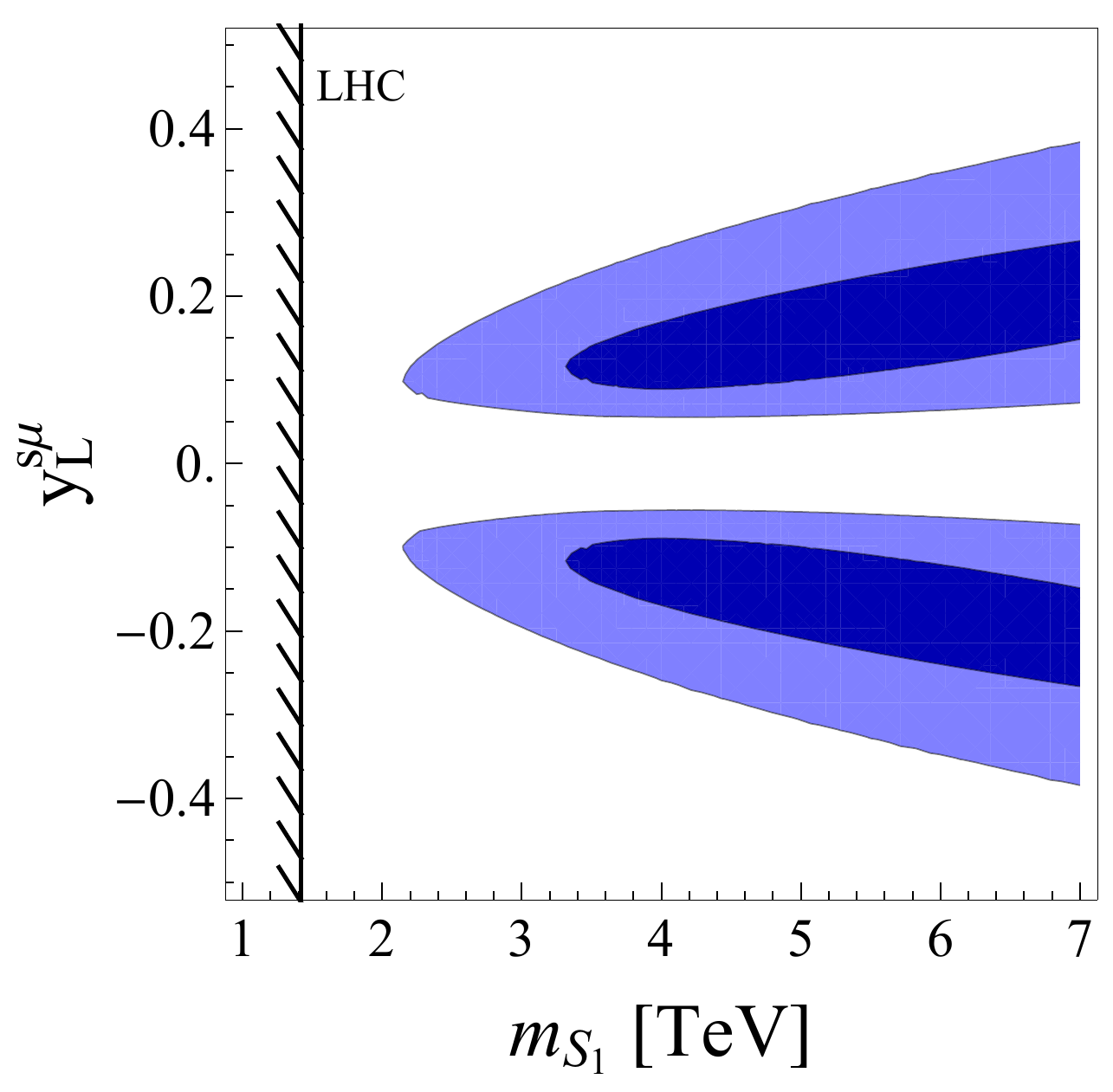}~\includegraphics[width=0.49\textwidth]{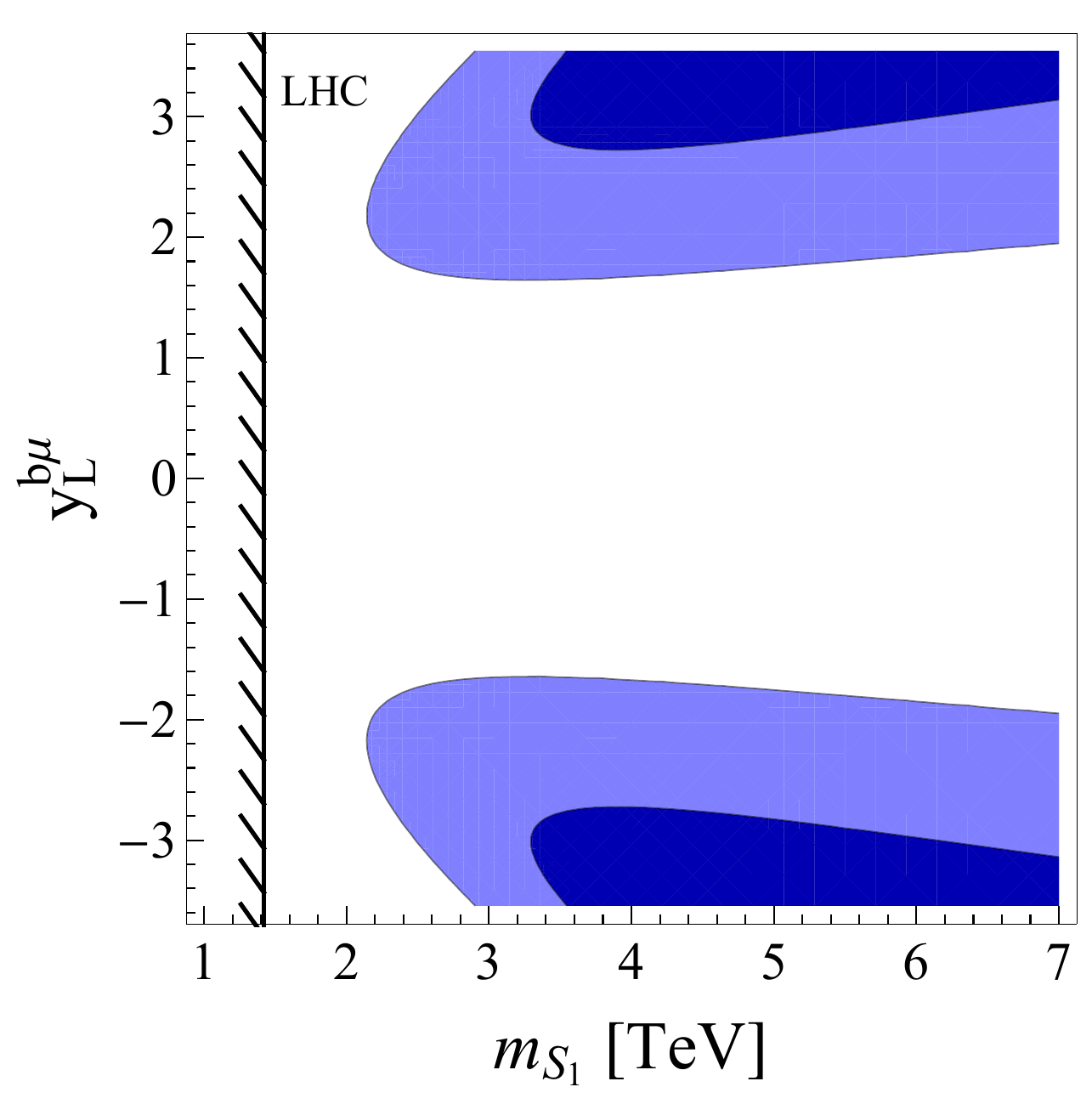}
  \caption{ \sl \small $m_{S_1}$ is plotted against $y_L^{s\mu}$ (left panel) and $y_L^{b\mu}$ (right panel). Light (dark) blue regions correspond to $\Delta \chi^2 <2.3 (6.8)$. The black line in the right panel denotes the LHC exclusion limit on pair produced LQs decaying into $t\mu$~\cite{CMS:2018itt}. See text for details.}
  \label{fig:coupl-S1-muons}
\end{figure}

So far we were concerned with the $b\to s\ell\ell$ anomalies, $R_{K^{(\ast)}}$. If instead we focus on $R_{D^{(\ast)}}$ then the flavor ansatz given in Eq.~\eqref{eq:xL-S1-pattern} must be extended to allow for $y_R^{c\tau}\neq 0$: 

\begin{equation}
\label{eq:xL-S1-pattern2}
y_L  = \begin{pmatrix}
0 & 0 & 0\\ 
0 & y_L^{s\mu} & y_L^{s\tau} \\ 
0 & y_L^{b\mu} & y_L^{b\tau}
\end{pmatrix}\,, \qquad\qquad y_R = \begin{pmatrix}
0 & 0 & 0\\ 
0 & 0 & y_R^{c\tau} \\ 
0 & 0 & y_R^{t\tau}
\end{pmatrix}\,.
\end{equation}

The couplings to $\tau$ bring in new constraints, such as those coming from $\mathcal{B}(D_s\to \tau \bar{\nu})$, $\mathcal{B}(B\to \tau \nu)$ and $\mathcal{B}(\tau \to \mu \gamma)$, as described in Ref.~\cite{Becirevic:2016oho}. By performing a $\chi^2$ analysis similar to the one above we conclude that the tension in $R_{K^{(\ast)}}$ and $R_{D^{(\ast)}}$ can be moderately reduced if one is to remain in the range of $m_{S_1}\lesssim 2$~TeV. To accommodate them one would need large couplings, $|y_L^{b\mu}|\approx \sqrt{4\pi}$ and $|y_R^{c\tau}| \approx \sqrt{4\pi}$, and $m_{S_1}\gtrsim 4$~TeV, which is different from the original proposal in Ref.~\cite{Bauer:2015knc}. These conclusions agree with Ref.~\cite{Cai:2017wry}.

	\item[•]$U_1=(\mathbf{3},\mathbf{1})_{2/3}$\\[0.4em] As discussed in Ref.~\cite{Buttazzo:2017ixm} the minimal $U_1$ model is one of the best candidates to simultaneously explain the anomalies in $R_{K^{(\ast)}}$ and $R_{D^{(\ast)}}$. The interesting features of this scenario are: (i) contributions to $R_{K^{(\ast)}}$ come from the Wilson coefficients $C_{9,10}$ satisfying $C_9=-C_{10}$,  (ii) a positive sign of $g_{V_L}$ allows to accommodate $R_{D^{(\ast)}}>R_{D^{(\ast)}}^\mathrm{SM}$, cf.~Eq.~\eqref{eq:gV-U1}, and (iii) the absence of tree-level contributions to $\mathcal{B}(B\to K^{(\ast)}\nu \bar{\nu})$, which is often a major obstacle to the models built to accommodate the $B$-physics anomalies. We will discuss this scenario in more detail in Sec.~\ref{sec:results}.
		
	\item[•]$U_3=(\mathbf{3},\mathbf{3})_{2/3}$\\[0.4em] Finally, the model $U_3$ is also a viable candidate to explain $R_{K^{(\ast)}}$, since it predicts $C_9=~-C_{10}$, but it cannot provide an explanation of $R_{D^{(\ast)}}> R_{D^{(\ast)}}^\mathrm{SM}$. This can be understood on the basis of Eq.~\eqref{eq:gV-U3} from which we learn that the leading contribution to $g_{V_L}$ comes with the wrong (negative) sign. The subleading couplings to the strange quark are constrained by the tight experimental limits on $\mathcal{B}(B\to K^{(\ast)} \nu \bar{\nu})$, which is why the net effect on $R_{D^{(\ast)}}$ is very small.
\end{itemize}

Our findings are summarized in Table~\ref{tab:LQ-lists}, from which we learn that $U_1$ is the only single LQ model that can simultaneously  accommodate $R_{K^{(\ast)}}$ and $R_{D^{(\ast)}}$, in agreement with findings of Ref.~\cite{Buttazzo:2017ixm}. A slightly non-minimalistic possibility is to build a model with two different scalar leptoquarks, as explored for $S_1$ and $S_3$ in Ref.~\cite{Buttazzo:2017ixm,Marzocca:2018wcf,Crivellin:2017zlb}, and for $R_2$ and $S_3$ in Ref.~\cite{Becirevic:2018afm}. Note that our conclusions can also serve as a guideline for future studies if one of the anomalies disappears. 

\

\begin{table}[t!]
\renewcommand{\arraystretch}{1.6}
\centering
\begin{tabular}{|c|c|c||c|}
\hline 
Model & $R_{K^{(\ast)}}$ & $R_{D^{(\ast)}}$ & $R_{K^{(\ast)}}$ $\&$ $R_{D^{(\ast)}}$\\ \hline\hline
$S_1$	& \,\,\color{red}\xmark$^{\color{blue}\ast}$ & $\color{blue}\checkmark$	& \,\,\color{red}\xmark$^{\color{blue}\ast}$	\\
$R_2$	&	\,\,\color{red}\xmark$^{\color{blue}\ast}$ & $\color{blue}\checkmark$	&\color{red}\xmark\\
$\widetilde{R_2}$	& \color{red}\xmark	&\color{red}\xmark	 &\color{red}\xmark	\\ 
$S_3$	& $\color{blue}\checkmark$	&\color{red}\xmark	 &\color{red}\xmark	\\   \hline
$U_1$	& $\color{blue}\checkmark$	& $\color{blue}\checkmark$	& $\color{blue}\checkmark$	\\
$U_3$	& $\color{blue}\checkmark$	&\color{red}\xmark	&\color{red}\xmark	\\  \hline
\end{tabular}
\caption{ \sl \small Summary of the LQ models which can accommodate $R_{K^{(\ast)}}$ (first column), $R_{D^{(\ast)}}$ (second column), and both $R_{K^{(\ast)}}$ and $R_{D^{(\ast)}}$ (third column) without inducing other phenomenological problems. The symbol \xmark$^{\ast}$ means that the discrepancy can be alleviated, but not fully accommodated.  See text for details.}
\label{tab:LQ-lists} 
\end{table}



\section{Revisiting $U_1=(\mathbf{{3}},\mathbf{1})_{2/3}$}
\label{sec:results}


In this Section we discuss in more detail the phenomenological status of the scenario $U_1$. 
We will use the low-energy physics observables which receive the tree-level contributions from the $U_1$ exchange to constrain the 
model parameters. We will also compare these results with the ones deduced from the experimental bounds based on direct searches at the LHC. 
Furthermore, we will make a brief comment concerning the loop effects.

		\subsection{Low-energy constraints}
		 
To satisfy both $R_{K^{(\ast)}} < R_{K^{(\ast)}}^\mathrm{SM}$ and $R_{D^{(\ast)}}> R_{D^{(\ast)}}^\mathrm{SM}$ we will assume the following structure for the Yukawa matrices:

\begin{equation}
\label{eq:yL-U1}
x_L  = \begin{pmatrix}
0 & 0 & 0\\ 
0 & x_L^{s\mu} & x_L^{s\tau}\\ 
0 & x_L^{b\mu} & x_L^{b\tau}
\end{pmatrix}\,, \qquad\qquad x_R = 0\,,
\end{equation}

\noindent where the couplings to the first generation are set to zero in order to avoid the conflicts with experimental limits on $\mu-e$ conversion on nuclei, the atomic parity violation and on $\mathcal{B}(K\to \pi \nu\bar{\nu})$. To determine the region allowed by $R_{K^{(\ast)}}^{\mathrm{exp}}$, we compare the result of the global fit to $b\to s\mu\mu$ observables, given in Eq.~\eqref{eq:C9-exp}, with the expression Eq.~\eqref{eq:C9-U1}, and find 

\begin{equation}
-\dfrac{x_L^{s\mu}\big{(}x_L^{b\mu}\big{)}^\ast}{m_{U_1}^2} \in [0.83,1.41]
\times 10^{-3}~\mathrm{TeV}^{-2}\,. 
\end{equation}

\noindent A tree-level contribution of $U_1$ to $R_{D^{(\ast)}}$ gives rise to the effective coefficient $g_{V_L}$, cf. Eq.~\eqref{eq:gV-U1}, which compared to Eq.~\eqref{eq:RDfit} results in 

\begin{equation}
\dfrac{\left(V x_L\right)_{c\tau}\left(x_L^{b\tau}\right)^\ast}{m_{U_1}^2}= \dfrac{(V_{cs}\, x_L^{s\tau}+V_{cb}\, x_L^{b\tau})\, (x_L^{b\tau})^\ast}{m_{U_1}^2} \in [0.12,0.18]~\mathrm{TeV}^{-2} \,.
\end{equation}
 
\noindent  Other relevant constraints to this scenario are listed in Table~\ref{tab:LQ-bounds}, including the decays $K\to \mu \bar{\nu}$, $D_{(s)}\to \tau\bar{\nu}$ and $B\to \tau\bar{\nu}$, as well as the ratio $R_D^{\mu/e}=\mathcal{B}(B\to D\mu\bar{\nu})/\mathcal{B}(B\to D e\bar{\nu})$. Another important constraint stems from $\mathcal{B}(\tau \to\mu\phi)^{\mathrm{exp}}<8.4 \times 10^{-8}$~\cite{Patrignani:2016xqp,Miyazaki:2011xe}, which was often neglected in previous studies of this particular LQ model, and which gives
(to $2\sigma$),

\begin{equation}
\label{eq:tau-muphi-U1}
\dfrac{|x_L^{s\mu}||x_L^{s\tau}|}{m_{U_1}^2} < 0.018~\mathrm{TeV}^{-2}\,.
\end{equation}

\noindent We reiterate that an important feature of this scenario is the absence of tree-level contributions to $B\to K^{(\ast)}\nu\bar{\nu}$. One-loop contributions to this transition as well as to many other observables, such as $\tau \to \mu\gamma$, $B_s-\bar{B}_s$ mixing and LFU tests in $\tau$ decays, can however be important. In Ref.~\cite{Feruglio:2016gvd} it was shown that the leading-log renormalization group effects induced by the effective operators related to $U_1$ could be in conflict with constraints from leptonic $Z$ and $\tau$ decays in the scenarios in which the dominant coupling is the one to the third generation fermions. To get around that difficulty one can allow for a more general flavor structure~\cite{Buttazzo:2017ixm}, similar to the one we consider here. In this work, we will not consider the constraints induced by the loop effects since they could be sensitive to the details of the unknown UV completion of the $U_1$ model, which is model dependent. We will simply assume that this problem is taken care of by some mechanism which prevents the appearances of divergences to higher order in perturbation theory. Moreover, as we shall see below, the synergy between the tree-level constraints from flavor physics and the tree-level bounds coming from direct searches at the LHC is already sufficient to significantly limit the parameter space of this scenario. Of course, this fact does not reduce importance of electroweak corrections which must be systematically included in the models in which the UV completion is specified.

\begin{table}[htbp!]
\renewcommand{\arraystretch}{1.5}
\centering
\begin{tabular}{|c|c|c|c|}
\hline 
Observable &  Eqs.~--~$U_1$  & Exp.~value & Ref. \\ \hline\hline
  $b\to s\mu\mu$ & \eqref{eq:C9-U1} & \eqref{eq:C9-exp}	& \cite{global} \\ 
  $b\to c\tau\nu$  &\eqref{eq:gV-U1}	&  \eqref{eq:RDfit} & \cite{Amhis:2016xyh} \\
  $\mathcal{B(\tau\to\mu\phi)}$ 	&\eqref{eq:tau-muphi-U1} &  $<8.4\times 10^{-8}$ & \cite{Patrignani:2016xqp}	\\
  $\mathcal{B}(B\to\tau\nu)$ 	&  \eqref{eq:gV-U1},~\eqref{eq:leptonicB}	& $1.06(19)\times 10^{-4}$  & \cite{Amhis:2016xyh}	\\
  $\mathcal{B}(D_s\to\mu\nu)$ 	& \eqref{eq:gV-U1},~\eqref{eq:leptonicB}	& $5.50(23)\times 10^{-3}$	&    \cite{Patrignani:2016xqp}	\\
  $\mathcal{B}(D_s\to\tau\nu)$ 	&  \eqref{eq:gV-U1},~\eqref{eq:leptonicB}	& $5.48(23)\times 10^{-2}$	  & \cite{Patrignani:2016xqp}	\\
  ${r_K^{e/\mu}}$ 	&	\eqref{eq:gV-U1},~\eqref{eq:leptonicB} &  $2.488(10)\times 10^{-5}$	&	\cite{Cirigliano:2007xi,Patrignani:2016xqp} \\
  ${r_K^{\tau/\mu}}$	& \eqref{eq:gV-U1},~\eqref{eq:leptonicB} & $4.670(67) \times 10^2$ & \cite{Dorsner:2017ufx}	\\
  ${R_D^{\mu/e}}$ 	& \eqref{eq:gV-U1}  &   $0.995(22)(39)$	& \cite{Glattauer:2015teq}	\\
 $\mathcal{B}(B\to K\mu\tau)$ &	 \eqref{eq:C9-U1} & $<4.8\times 10^{-5}$ & \cite{Lees:2012zz} \\
  \hline
\end{tabular}
\caption{ \sl \small Tree-level observables considered in our phenomenological analysis and their corresponding experimental values (or limits), as well as their theoretical expressions for the $U_1$ scenario.}
\label{tab:LQ-bounds} 
\end{table}
	
\subsection{Results and predictions}

The results of our analysis will be presented in two parts. In the first we set $m_{U_1} = 1.5$~TeV, which is the lowest $U_1$ mass not yet excluded by vector LQ pair production searches at the LHC~\cite{CMS:2018bhq}. The resulting parameter space will then be used to show our predictions for two LFV processes, $B \to K \mu \tau$ and $\tau \to \mu \phi$. 
In the second part, we repeat the same exercise but this time treating $m_{U_1}$ as a free parameter.
 
\subsubsection*{Scan of parameters with $m_{U_1}=1.5$~TeV}

For our analysis with fixed $m_{U_1} = 1.5$~TeV, we first find a best fit point by minimizing a $\chi^2$-function built from the flavor observables listed in Table~\ref{tab:LQ-bounds}. We find $\chi^2_{\rm min} = 6.61$ for
\begin{align}
\label{eq:chiSQmin_U1}
x_L^{s\mu} \approx -10^{-2}\,, \qquad x_L^{b\mu} \approx 0.25\,, \qquad x_L^{s\tau} \approx 4.4\times 10^{-3}\,, \qquad x_L^{b\tau} \approx 2.81\,.
\end{align} 

\noindent We then perform a random scan over the values of the four left-handed couplings shown in Eq.~\eqref{eq:yL-U1}, and enforce perturbativity, $|x^{ij}_L| \leq \sqrt{4\pi}$. We select only the points which satisfy $\Delta \chi^2 ({\rm par}) \equiv \chi^2 ({\rm par}) - \chi^2_{\rm min} \leq 6.18 $, i.e. within $2\,\sigma$ from the best fit point. The selected points are compared with the limits deduced from the direct LHC searches in $\ell \ell$ ($\ell$ =$\mu$, $\tau$) final states, as detailed in Sec.~\ref{sec:lhc-dilepton}. In the plots presented in this Section, the points excluded by direct searches based on current LHC data (36~fb$^{-1}$) are shown in grey. Furthermore, the red points are those that are excluded from our projections to 300~fb$^{-1}$. The blue points are those that would survive. 

\begin{figure}[h!]
  \centering
  \includegraphics[width=0.6\textwidth]{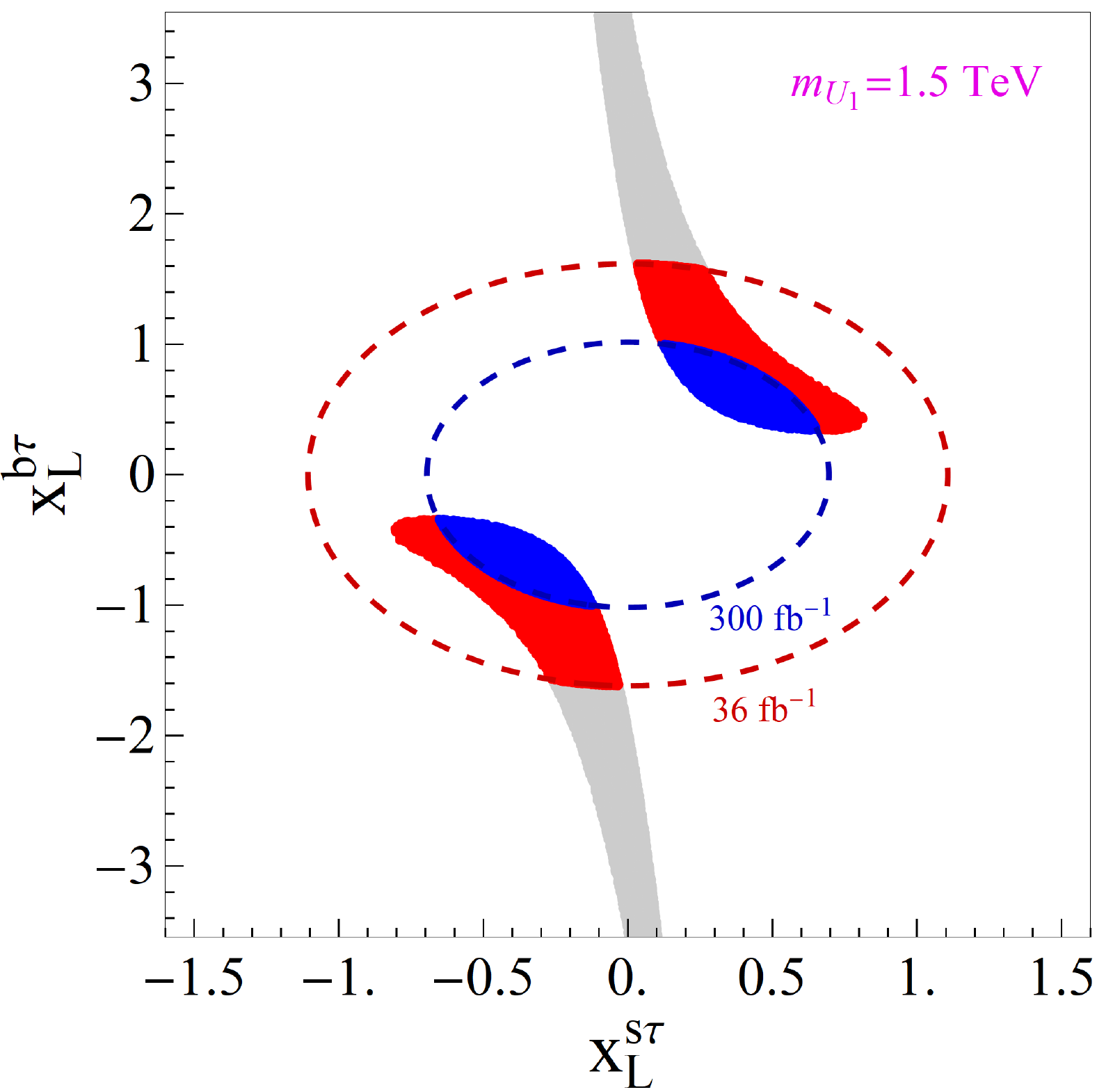}
  \caption{ \sl \small The correlation between the couplings $x_L^{s\tau}$ and $x_L^{b\tau}$ allowed by flavor constraints is plotted for $m_{U_1}=1.5$~TeV. Gray points are excluded by current LHC data ($36~\mathrm{fb}^{-1}$) on $pp\to \ell\ell$ ($\ell=\mu,\tau$). The future LHC sensitivity is depicted by the red points, which were obtained by extrapolating current data to $300~\mathrm{fb}^{-1}$, as discussed in Sec.~\ref{sec:lhc-dilepton}. Blue points are allowed by all constraints, including the extrapolated LHC results to $300~\mathrm{fb}^{-1}$.}
  \label{fig:couplings-tau-U1}
\end{figure}

We first show, in Fig.~\ref{fig:couplings-tau-U1}, the correlation between the two LQ couplings entering Eq.~\eqref{eq:gV-U1} for $m_{U_1}=1.5$~TeV. One observes that the experimental value of $R_{D^{(\star)}}$ forces $x_L^{b\tau}$ to be different from zero, thus bounding its absolute value from below. Even though the measurements of low-energy observables allow for $x_L^{s\tau}=0$, we see that current LHC data exclude this possibility, imposing the lower bound $|x_L^{s\tau}|\gtrsim 0.03$. Moreover, our projected bound for 300~fb$^{-1}$ will push this limit even further away from $0$, implying $|x_L^{s\tau}|\gtrsim 0.1$. Even though we opted to not consider the model dependent radiative bounds, which are derived within the leading logarithm approximation~\cite{Feruglio:2016gvd}, it is interesting to note that comparable lower bounds on $|x_L^{s\tau}|$ can be obtained by these means~\cite{Buttazzo:2017ixm}.

\begin{figure}[htbp!]
  \centering
  \includegraphics[width=0.5\textwidth]{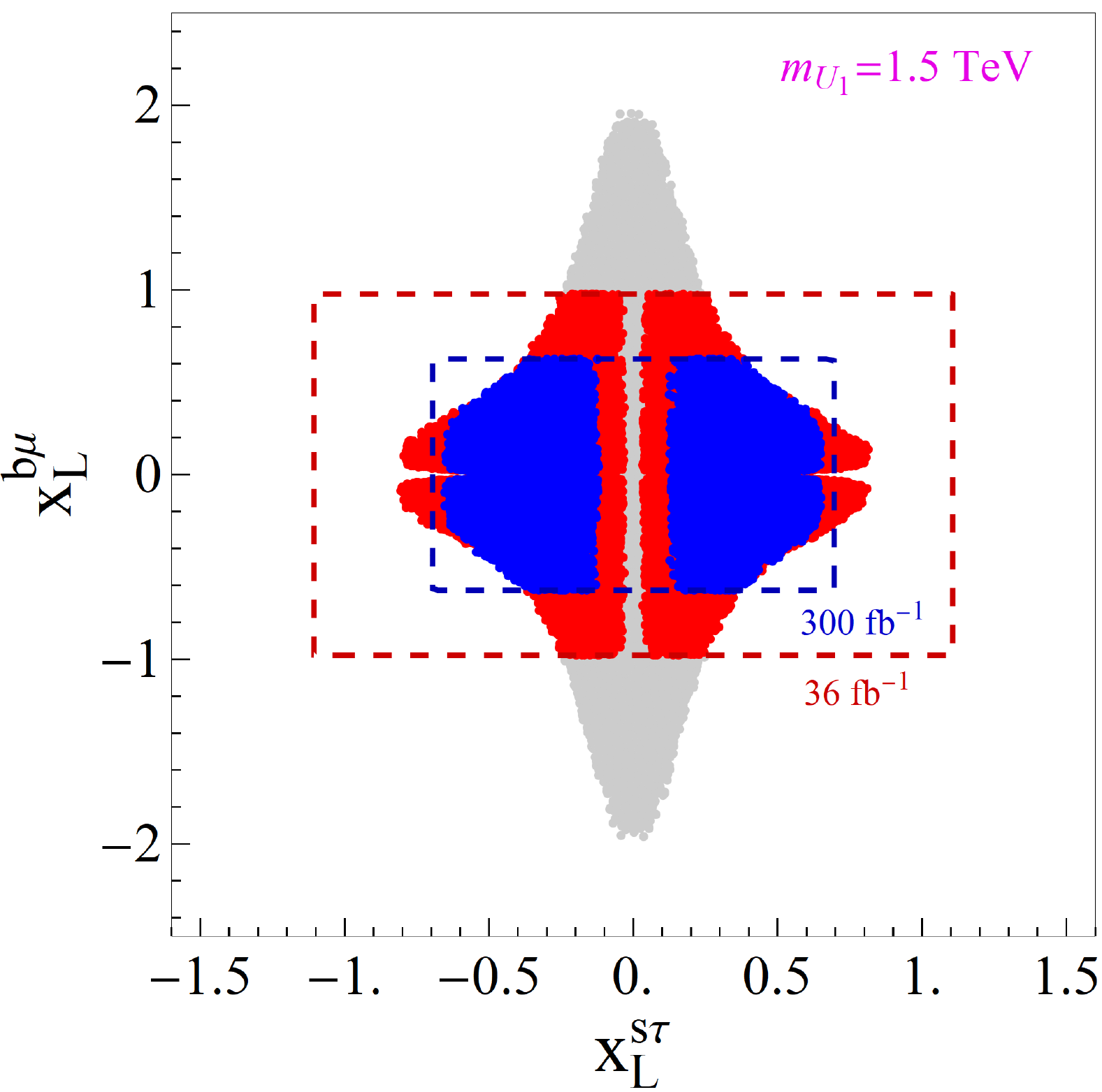}~\includegraphics[width=0.5\textwidth]{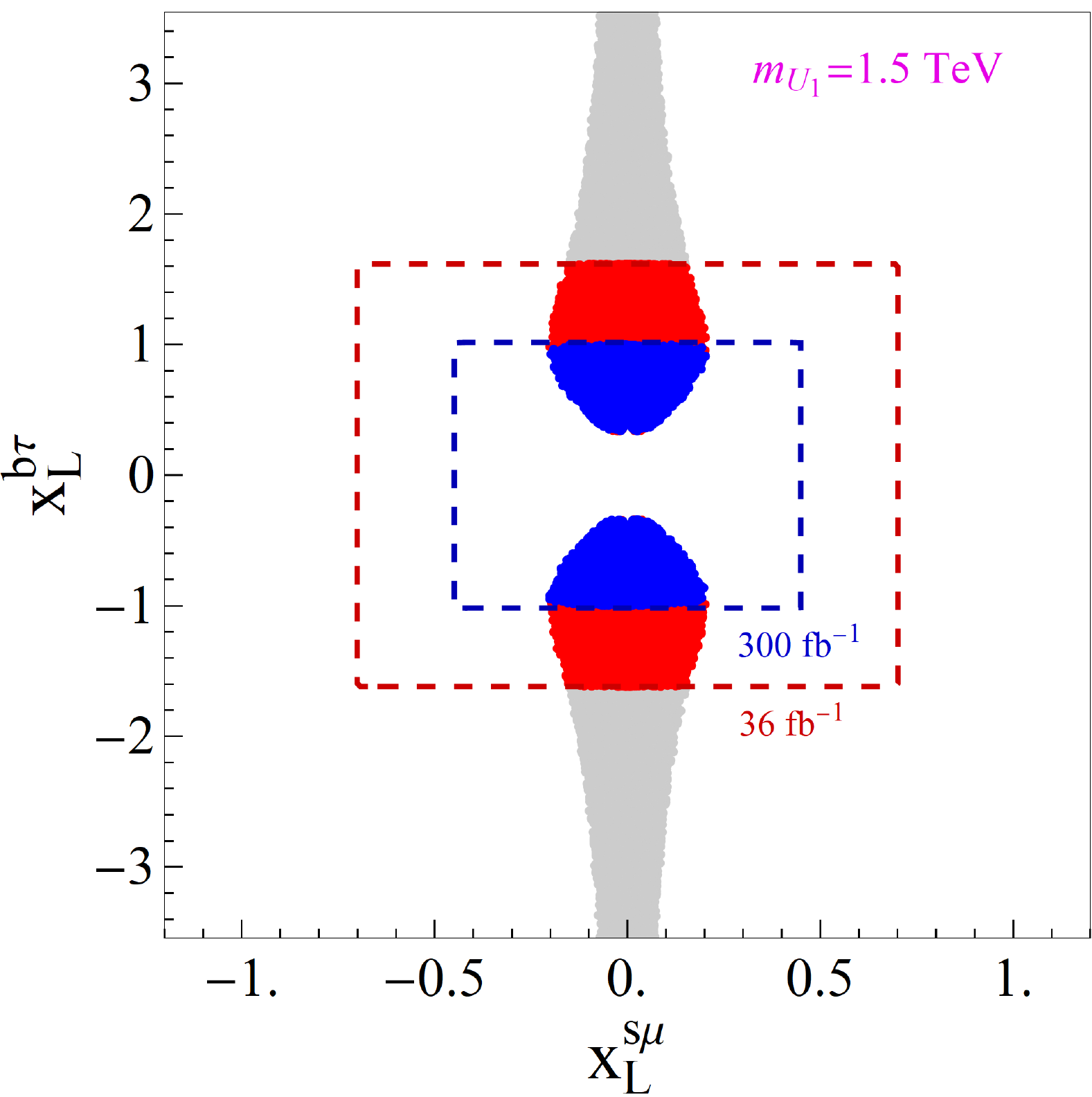}   
  \caption{ \sl \small Coupling $x_L^{s\tau}$ is plotted against $x_L^{b\mu}$ (left panel), and  $x_L^{b\tau}$ is plotted against $x_L^{s\mu}$ (right panel), by assuming $m_{U_1}=1.5$~TeV. Color code is the same as in Fig.~\ref{fig:couplings-tau-U1}. }
  \label{fig:couplings-tau-mu-U1}
\end{figure}

Next, we show in Fig.~\ref{fig:couplings-tau-mu-U1} the correlations between $x_L^{b\mu}$ and $x_L^{s\tau}$ (left panel) and between $x_L^{b\tau}$ and $x_L^{s\mu}$ (right panel). The color code remains the same as before and the red (blue) dashed lines correspond to the LHC limits obtained at $36$~fb$^{-1}$ ($300$~fb$^{-1}$), but assuming for simplicity that the couplings which are not present in a given plot are set to zero.~\footnote{Setting other couplings to zero to get the dashed regions in these plots is the reason why some of the red points remain within the dashed blue rectangles (because for these points the other couplings which are not in the plot are not set to zero).}  It is clear from Eq.~\eqref{eq:C9-U1} that in order to explain the measured deviation with respect to the SM in the $b\to s \mu \mu$ transitions, both $x_L^{s\mu}$ and $x_L^{b\mu}$ need to be different from zero. Moreover, as discussed in the case of Fig.~\ref{fig:couplings-tau-U1}, current and future LHC limits provide a lower bound on $|x_L^{s\tau}|$, while $R_{D^{(\star)}}^{\rm exp}$ sets a lower limit on $|x_L^{b\tau}|$. These considerations have an important impact on the LFV decays, as we discuss below.

\begin{figure}[htbp!]
  \centering
 \includegraphics[width=0.65\textwidth]{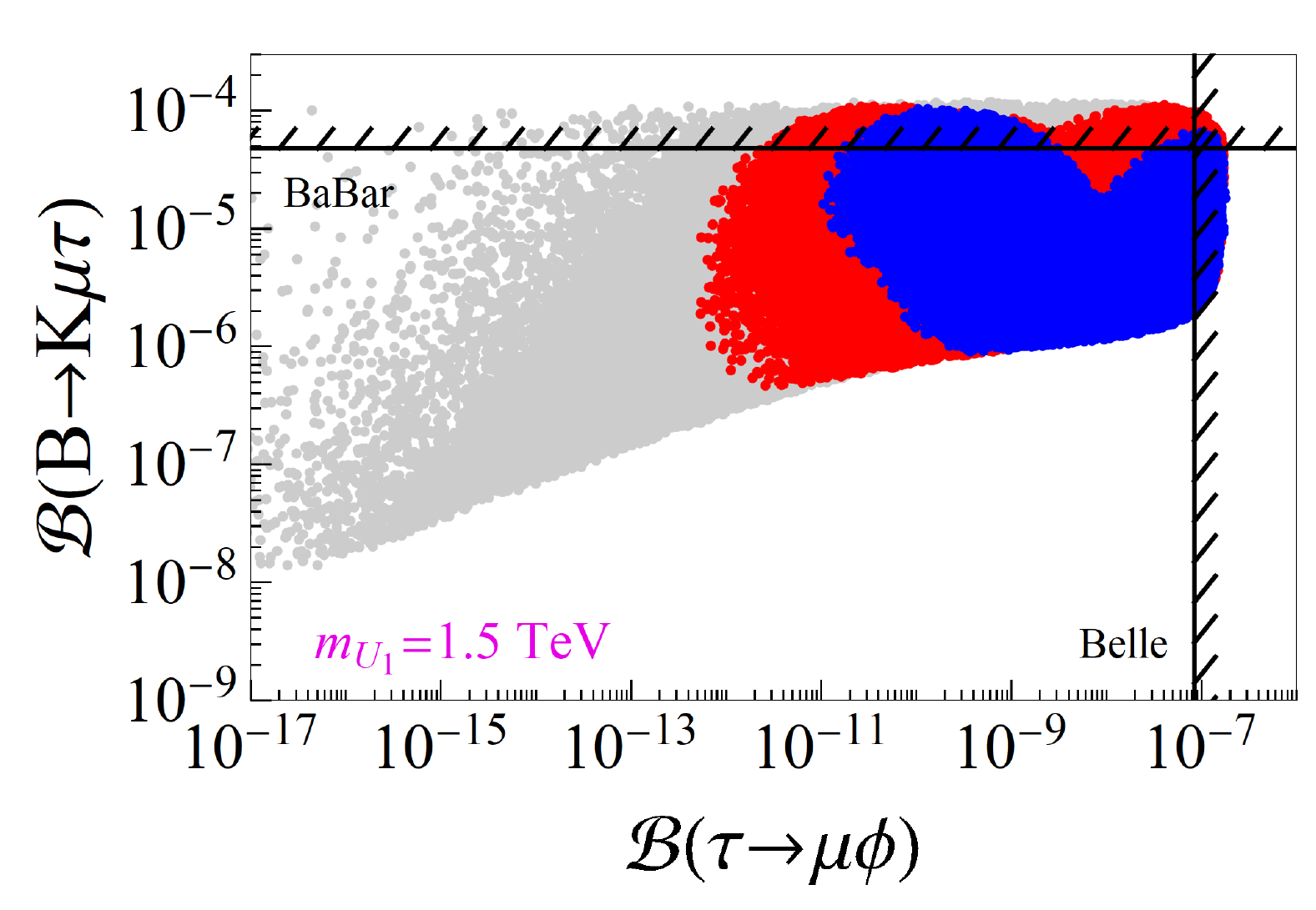}
  \caption{ \sl \small $\mathcal{B}(B\to K\mu\tau)$ is plotted against $\mathcal{B}(\tau\to\mu\phi)$ for the $U_1$ model. Color code is the same as in Fig.~\ref{fig:couplings-tau-U1}. Current bounds on these two decays, as respectively established by BaBar~\cite{Lees:2012zz} and by Belle~\cite{Miyazaki:2011xe}, are also shown.}
  \label{fig:correlation-U1}
\end{figure}

 We finally show in Fig.~\ref{fig:correlation-U1} our prediction for the correlation of two LFV observables, $\mathcal{B}(\tau \to \mu \phi)$ and $\mathcal{B}(B \to K \mu \tau)$, with the hatched black lines denoting the current experimental bounds on these processes. Again, $m_{U_1}$ is set to 1.5~TeV. As mentioned in the previous paragraph, the fact that the  
LHC sets a lower bound on the absolute value of $|x_L^{s\tau}|$ has a dramatic impact on the amount of LFV predicted by the $U_1$ model: as can be seen in Fig.~\ref{fig:correlation-U1}. Interestingly, we see that the current LHC bounds lead to $\mathcal{B}(B \to K \mu \tau)\gtrsim 2\times 10^{-7}$, which remains rather stable lower bound for the LFV mode. With $300$~\invfb we get that this bound is improved to $\mathcal{B}(B \to K \mu \tau)\gtrsim 5\times 10^{-7}$. In other words, we get an absolute lower bound of $\mathcal{O}(10^{-7})$.  We see that lowering the upper bound on $\mathcal{B}(B \to K \mu \tau)$ at the LHCb and/or Belle~II can have a major impact on the model building by further restraining the parameter space. 

\subsubsection*{Scan of parameters with varying $m_{U_1}$}

\begin{figure}[t!]
  \centering
 \includegraphics[width=0.5\textwidth]{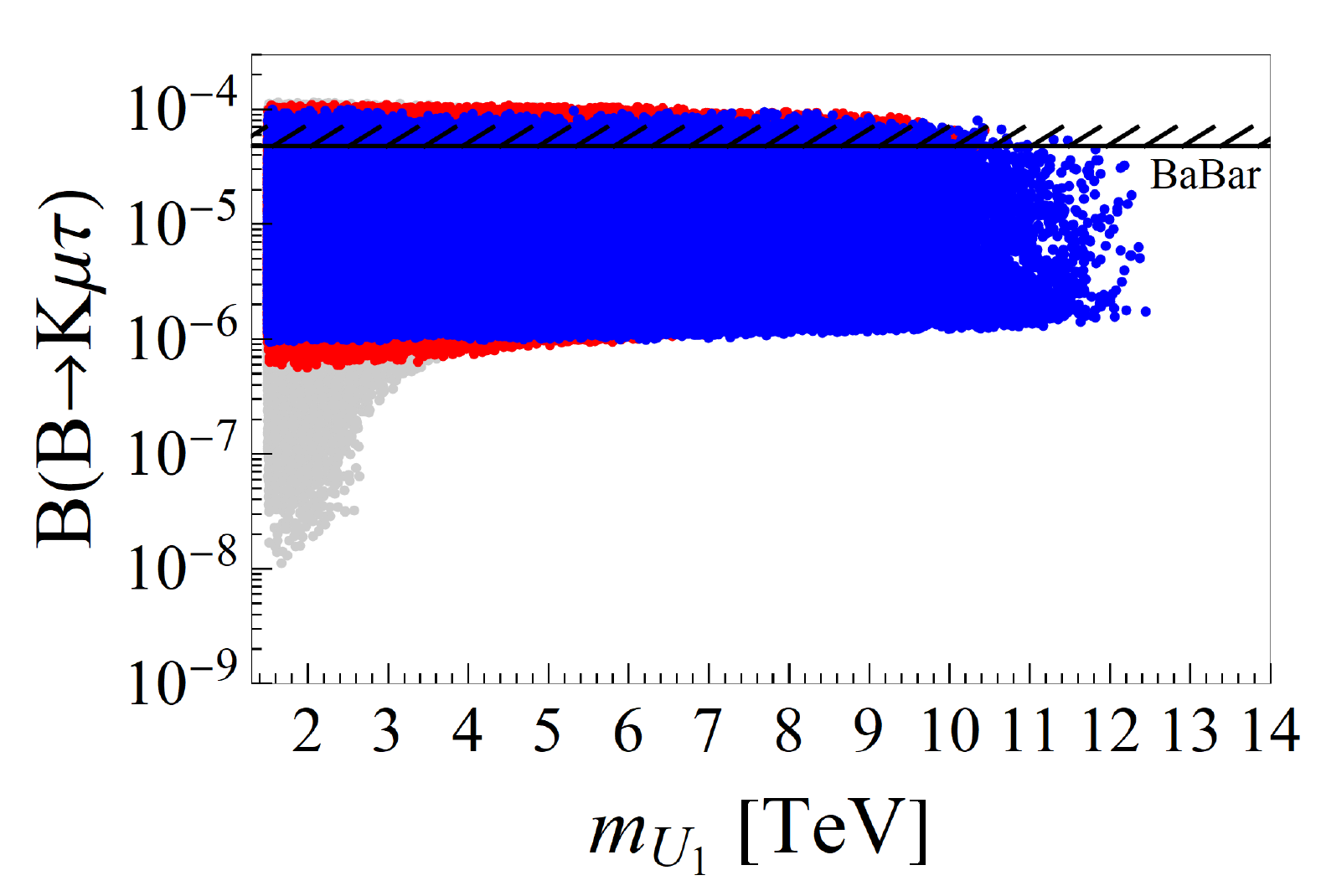}~\includegraphics[width=0.5\textwidth]{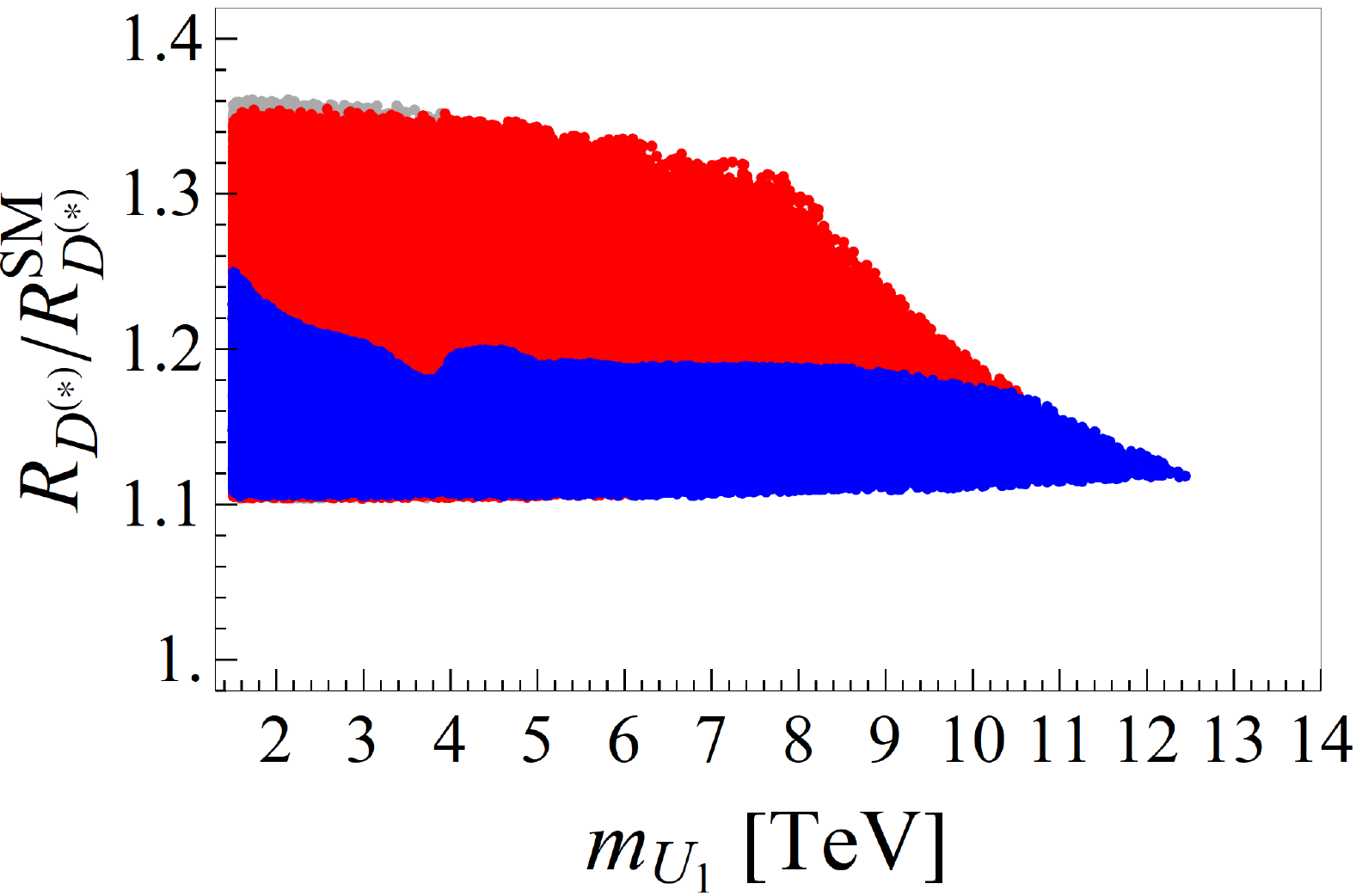}
  \caption{ \sl \small $m_{U_1}$ is plotted against $\mathcal{B}(B\to K\mu\tau)$ (left panel) and $R_{D^{(\ast)}}/R_{D^{(\ast)}}^{\mathrm{SM}}$ (right panel) for the $U_1$ model. Color code is the same as in Fig.~\ref{fig:couplings-tau-U1}.}
  \label{fig:correlation-U1_floatmass}
\end{figure}

 We now repeat the same analysis as before but by letting $m_{U_1}$ to be a free parameter too. $\chi^2_\mathrm{min}$ corresponds to the same couplings given in  Eq.~\eqref{eq:chiSQmin_U1}, except that now $\chi^2$ is minimized along a line in parameter space defined by a constant $x_L^{ij}/m_{U_1}$, with $|x_L^{ij}| < \sqrt{4\pi}$ (perturbativity limit).~\footnote{This scale invariance relationship holds in the case of $U_1$ because we only consider tree-level flavor constraints, and all the tree-level Wilson coefficients entering the analysis scale as (coupling/mass)$^2$.}  The main result of that analysis is shown in Fig.~\ref{fig:correlation-U1_floatmass}, where we show how the LFV branching fraction $\mathcal{B}(B\to K\mu\tau)$ (left panel) and $R_{D^{(\ast)}}/R_{D^{(\ast)}}^{\mathrm{SM}}$ (right panel) depend on $m_{U_1}$. We notice that, in order to be able to explain both flavor anomalies within the $U_1$ framework, $m_{U_1}$ cannot be higher than $\sim 12.5$~TeV.~\footnote{Note that this value is similar to the upper bound derived from unitarity of $2\to 2$ fermion scattering amplitudes, namely~$\Lambda \approx 9$~TeV for the $b\to c$ transition~\cite{DiLuzio:2017chi} .} This upper bound roughly corresponds to setting $x_L^{b\tau} = x_L^{s\tau} = \sqrt{4\pi}$ and looking for the highest value of the mass for which the $b \to c \tau \nu$ anomaly can still be explained within $2\,\sigma$.  Moreover the lower bounds on both quantities remain rather stable when varying $m_{U_1}$.
 
Finally, the plots presented in Fig.~\ref{fig:correlation-U1_floatmass} also showcase the complementarity between flavor physics (indirect) and LHC high-$p_T$ dilepton (direct) searches in constraining the $U_1$ scenario as a possible explanation of the flavor anomalies. While accommodating $R_{K^{(\ast)}} < R_{K^{(\ast)}}^\mathrm{SM}$ and $R_{D^{(\ast)}}> R_{D^{(\ast)}}^\mathrm{SM}$ already results in an absolute lower bound $\mathcal{B}(B\to K\mu\tau) \gtrsim 10^{-8}$, that bound gets shifted upwards by an order of magnitude to $\sim 5 \times 10^{-7}$ if one accounts for the direct LHC searches with $36\ \mathrm{fb}^{-1}$ of data and for $m_{U_1} < 4$~TeV. Going to projected $300\ \mathrm{fb}^{-1}$ leaves the lower bound on both $\mathcal{B}(B\to K\mu\tau)$ and  $R_{D^{(\ast)}}/R_{D^{(\ast)}}^\mathrm{SM}$ quite stable. The upper bound on $\mathcal{B}(B\to K\mu\tau)$ is also stable but already superseded by the experimental limit on this decay mode established by BaBar. Measuring (bounding) $\mathcal{B}(B\to K\mu\tau)$ thus becomes very appealing as the improvement of the current upper bound can either help discarding the $U_1$ scenario altogether, or further corroborate its viability. 
Note that for definiteness we focus on $B\to K\mu\tau$, but the discussion would be completely equivalent  if we discussed $B_s\to \mu\tau$ or $B\to K^\ast \mu\tau$, because their branching fractions are known to be related via~\cite{Becirevic:2016zri} 
\bea
{\mathcal{B}(B\to K^\ast \mu\tau)\over \mathcal{B}(B\to K\mu\tau)}\approx 1.8\,,\qquad\qquad {\mathcal{B}(B_s\to \mu\tau)\over \mathcal{B}(B\to K\mu\tau)}\approx 0.9\,.
\eea
This kind of LFV decay modes was also mentioned in Ref.~\cite{Bordone:2018nbg} as a good probe of validity of a specific UV completion of the low-energy $U_1$ model. 

As for the upper limit on $R_{D^{(\ast)}}/R_{D^{(\ast)}}^\mathrm{SM}$, we see that the direct searches can play a very important role in further reducing the space of parameters and with $300\ \mathrm{fb}^{-1}$ of data at the LHC the possible range of values for this ratio reduces to $1.1\leq R_{D^{(\ast)}}/R_{D^{(\ast)}}^\mathrm{SM} \leq 1.25$. 

\section{Summary and conclusion}
\label{sec:conclusions}

In this work we revisited the single LQ solutions to the $B$-physics anomalies, $R_{D^{(\ast)}}^\mathrm{exp}> R_{D^{(\ast)}}^\mathrm{SM}$ and/or $R_{K^{(\ast)}}^\mathrm{exp}< R_{K^{(\ast)}}^\mathrm{SM}$. We find that none of the scalar LQs alone, with the mass $m_\mathrm{LQ}\simeq 1$~TeV, can provide a model of NP that accommodates simultaneously both kinds of anomalies. To arrive to that conclusion we combined a number of constraints on the model parameters arising from the low-energy flavor physics observables with those coming from the direct searches at the LHC. Concerning the latter ones the most significant constraints come from the large-$p_T$ spectrum of the differential cross section of $pp\to \ell\ell$. We use the most recent experimental results which we recast to obtain the bounds relevant to each of the models considered in this work. Since none of the scalar LQs can alone satisfy all the constraints (including the $B$-physics anomalies), see Table~\ref{tab:LQ-lists}, a promising route for model building involving leptoquarks seems to be  combining two scalar LQs, in a way that has been done in Refs.~\cite{Becirevic:2018afm,Marzocca:2018wcf,Crivellin:2017zlb}. 

Besides the scalar LQs we also considered the vector ones. The main difficulty in this case is that one has to specify a UV completion of the model in order to compute the loop effects.
By focusing only on the tree level observables, we confirm that the weak singlet vector LQ ($U_1$) of mass $m_\mathrm{LQ}\simeq 1\div 2$~TeV can indeed accommodate both $R_{D^{(\ast)}}^\mathrm{exp}> R_{D^{(\ast)}}^\mathrm{SM}$ and $R_{K^{(\ast)}}^\mathrm{exp}< R_{K^{(\ast)}}^\mathrm{SM}$, in its minimal version, i.e. by allowing non-zero values only to the left-handed couplings~\cite{Buttazzo:2017ixm}.  We find that the new results from direct searches indeed push the lower bound of the vector LQ to larger values (above $1.5~\tev$), and observe a pronounced complementarity of the low-energy (flavor physics) constraints with those obtained from direct searches. In particular we find the lower bound on the LFV mode 
$\mathcal{B}(B\to K\mu\tau) \gtrsim \mathrm{few} \times 10^{-7}$ for any mass of $m_{U_1}$ in which Yukawa couplings are kept within the perturbativity limits and in the minimal $U_1$ scenario in which only left-handed couplings are allowed to take values different from zero. 
Notice that the upper bound is superseded by the current experimental bound $\mathcal{B}(B\to K\mu\tau)^\mathrm{exp} < 4.8\times 10^{-5}$, which can be improved both at 
LHCb and Belle~II. Improving that bound by two orders of magnitude can therefore either exclude or, if observed, corroborate the validity of the minimal $U_1$ scenario.

\section{Acknowledgments}
\label{sec:acknowledgments}

This project has received support from the European Union's Horizon 2020 research and innovation programme under the Marie Sklodowska-Curie grant agreement N$^\circ$~674896 and N$^\circ$~690575. Work at University of Nebraska-Lincoln is supported by the Department of Physics and Astronomy. A.A. would like to thank Frank Golf for insightful discussions on leptoquark searches at the LHC. D.A.F is  supported  by  the {\it Young Researchers Programme} of the Slovenian Research  Agency under the grant N$^\circ$~37468.


\begin{thebibliography}{99}

\bibitem{Lees:2012xj} 
  J.~P.~Lees {\it et al.} [BaBar Collaboration],
  Phys.\ Rev.\ Lett.\  {\bf 109}, 101802 (2012)
  [arXiv:1205.5442 [hep-ex]].


\bibitem{Lees:2013uzd} 
  J.~P.~Lees {\it et al.} [BaBar Collaboration],
  Phys.\ Rev.\ D {\bf 88}, no. 7, 072012 (2013)
  [arXiv:1303.0571 [hep-ex]].


\bibitem{Huschle:2015rga} 
  M.~Huschle {\it et al.} [Belle Collaboration],
  Phys.\ Rev.\ D {\bf 92}, no. 7, 072014 (2015)
  [arXiv:1507.03233 [hep-ex]].


\bibitem{Aaij:2015yra} 
  R.~Aaij {\it et al.} [LHCb Collaboration],
  Phys.\ Rev.\ Lett.\  {\bf 115}, no. 11, 111803 (2015)
  Erratum: [Phys.\ Rev.\ Lett.\  {\bf 115}, no. 15, 159901 (2015)]
  [arXiv:1506.08614 [hep-ex]].


\bibitem{Hirose:2016wfn} 
  S.~Hirose {\it et al.} [Belle Collaboration],
  Phys.\ Rev.\ Lett.\  {\bf 118}, no. 21, 211801 (2017)
  [arXiv:1612.00529 [hep-ex]].


\bibitem{Sato:2016svk} 
  Y.~Sato {\it et al.} [Belle Collaboration],
  Phys.\ Rev.\ D {\bf 94}, no. 7, 072007 (2016)
  [arXiv:1607.07923 [hep-ex]].


\bibitem{Abdesselam:2016cgx} 
  A.~Abdesselam {\it et al.} [Belle Collaboration],
  arXiv:1603.06711 [hep-ex].



\bibitem{Amhis:2016xyh} 
  Y.~Amhis {\it et al.} [HFLAV Collaboration],
  Eur.\ Phys.\ J.\ C {\bf 77}, no. 12, 895 (2017)
  [arXiv:1612.07233 [hep-ex]], for regular updates please see https://hflav.web.cern.ch/content/semileptonic-b-decays .





\bibitem{Lattice:2015rga} 
  J.~A.~Bailey {\it et al.} [MILC Collaboration],
  Phys.\ Rev.\ D {\bf 92}, no. 3, 034506 (2015)
  [arXiv:1503.07237 [hep-lat]].


\bibitem{Na:2015kha} 
  H.~Na {\it et al.} [HPQCD Collaboration],
  Phys.\ Rev.\ D {\bf 92}, no. 5, 054510 (2015)
  Erratum: [Phys.\ Rev.\ D {\bf 93}, no. 11, 119906 (2016)]
  [arXiv:1505.03925 [hep-lat]].

\bibitem{Aoki:2016frl}
  S.~Aoki {\it et al.},
  Eur.\ Phys.\ J.\ C {\bf 77} (2017) no.2,  112
  [arXiv:1607.00299 [hep-lat]].


\bibitem{Bigi:2017jbd}
  D.~Bigi, P.~Gambino and S.~Schacht,
  JHEP {\bf 1711} (2017) 061
  [arXiv:1707.09509 [hep-ph]];
  S.~Jaiswal, S.~Nandi and S.~K.~Patra,
  JHEP {\bf 1712} (2017) 060
  [arXiv:1707.09977 [hep-ph]].


\bibitem{Bernlochner:2017jka} 
  F.~U.~Bernlochner, Z.~Ligeti, M.~Papucci and D.~J.~Robinson,
  Phys.\ Rev.\ D {\bf 95}, no. 11, 115008 (2017)
  Erratum: [Phys.\ Rev.\ D {\bf 97}, no. 5, 059902 (2018)]
  [arXiv:1703.05330 [hep-ph]].


\bibitem{Aaij:2017tyk}
  R.~Aaij {\it et al.} [LHCb Collaboration],
  Phys.\ Rev.\ Lett.\  {\bf 120} (2018) no.12,  121801
  [arXiv:1711.05623 [hep-ex]].





\bibitem{Aaij:2014ora} 
  R.~Aaij {\it et al.} [LHCb Collaboration],
  Phys.\ Rev.\ Lett.\  {\bf 113}, 151601 (2014)
  [arXiv:1406.6482 [hep-ex]].


\bibitem{Aaij:2017vbb} 
  R.~Aaij {\it et al.} [LHCb Collaboration],
  JHEP {\bf 1708}, 055 (2017)
  [arXiv:1705.05802 [hep-ex]].


\bibitem{Bordone:2016gaq}
  M.~Bordone, G.~Isidori and A.~Pattori,
  Eur.\ Phys.\ J.\ C {\bf 76} (2016) no.8,  440
  [arXiv:1605.07633 [hep-ph]];
G.~Hiller and F.~Kruger,
  Phys.\ Rev.\ D {\bf 69} (2004) 074020
  [hep-ph/0310219].


\bibitem{Becirevic:2016oho} 
  D.~Bečirević, N.~Košnik, O.~Sumensari and R.~Zukanovich Funchal,
  JHEP {\bf 1611}, 035 (2016)
  [arXiv:1608.07583 [hep-ph]].


\bibitem{Altmannshofer:2017poe}
  W.~Altmannshofer, P.~S.~Bhupal Dev and A.~Soni,
  Phys.\ Rev.\ D {\bf 96} (2017) no.9,  095010
  [arXiv:1704.06659 [hep-ph]].


\bibitem{Glashow:2014iga}
  S.~L.~Glashow, D.~Guadagnoli and K.~Lane,
  Phys.\ Rev.\ Lett.\  {\bf 114} (2015) 091801
  [arXiv:1411.0565 [hep-ph]].



\bibitem{Bobeth:1999mk}
  C.~Bobeth, M.~Misiak and J.~Urban,
  Nucl.\ Phys.\ B {\bf 574} (2000) 291
  [hep-ph/9910220].




\bibitem{Becirevic:2016zri} 
  D.~Bečirević, O.~Sumensari and R.~Zukanovich Funchal,
  Eur.\ Phys.\ J.\ C {\bf 76}, no. 3, 134 (2016)
  [arXiv:1602.00881 [hep-ph]].


\bibitem{global} 
  B.~Capdevila, A.~Crivellin, S.~Descotes-Genon, J.~Matias and J.~Virto,
  JHEP {\bf 1801}, 093 (2018)
  [arXiv:1704.05340 [hep-ph]];
  G.~D'Amico, M.~Nardecchia, P.~Panci, F.~Sannino, A.~Strumia, R.~Torre and A.~Urbano,
  JHEP {\bf 1709}, 010 (2017)
  [arXiv:1704.05438 [hep-ph]];
  W.~Altmannshofer, P.~Stangl and D.~M.~Straub,
  Phys.\ Rev.\ D {\bf 96}, no. 5, 055008 (2017)
  [arXiv:1704.05435 [hep-ph]];
  T.~Hurth, F.~Mahmoudi, D.~Martinez Santos and S.~Neshatpour,
  Phys.\ Rev.\ D {\bf 96}, no. 9, 095034 (2017)
  [arXiv:1705.06274 [hep-ph]].


\bibitem{Aaij:2017vad} 
  R.~Aaij {\it et al.} [LHCb Collaboration],
  Phys.\ Rev.\ Lett.\  {\bf 118}, no. 19, 191801 (2017)
  [arXiv:1703.05747 [hep-ex]].


\bibitem{Bobeth:2013uxa} 
  C.~Bobeth, M.~Gorbahn, T.~Hermann, M.~Misiak, E.~Stamou and M.~Steinhauser,
  Phys.\ Rev.\ Lett.\  {\bf 112}, 101801 (2014)
  [arXiv:1311.0903 [hep-ph]].



\bibitem{Dorsner:2016wpm} 
  I.~Doršner, S.~Fajfer, A.~Greljo, J.~F.~Kamenik and N.~Košnik,
  Phys.\ Rept.\  {\bf 641}, 1 (2016)
  [arXiv:1603.04993 [hep-ph]].



\bibitem{Aebischer:2015fzz} 
  J.~Aebischer, A.~Crivellin, M.~Fael and C.~Greub,
  JHEP {\bf 1605}, 037 (2016)
  [arXiv:1512.02830 [hep-ph]].



\bibitem{Jung:2018lfu} 
  M.~Jung and D.~M.~Straub,
  arXiv:1801.01112 [hep-ph].



\bibitem{Freytsis:2015qca} 
  M.~Freytsis, Z.~Ligeti and J.~T.~Ruderman,
  Phys.\ Rev.\ D {\bf 92}, no. 5, 054018 (2015)
  [arXiv:1506.08896 [hep-ph]].




\bibitem{Becirevic:2018uab} 
  D.~Bečirević, B.~Panes, O.~Sumensari and R.~Zukanovich Funchal,
  JHEP {\bf 1806}, 032 (2018)
  [arXiv:1803.10112 [hep-ph]].


\bibitem{Becirevic:2012jf}
  D.~Becirevic, N.~Kosnik and A.~Tayduganov,
  Phys.\ Lett.\ B {\bf 716} (2012) 208
  [arXiv:1206.4977 [hep-ph]];
  A.~Celis, M.~Jung, X.~Q.~Li and A.~Pich,
  Phys.\ Lett.\ B {\bf 771} (2017) 168
  [arXiv:1612.07757 [hep-ph]];
  JHEP {\bf 1301} (2013) 054
  [arXiv:1210.8443 [hep-ph]];
  P.~Biancofiore, P.~Colangelo and F.~De Fazio,
  Phys.\ Rev.\ D {\bf 87} (2013) no.7,  074010
  [arXiv:1302.1042 [hep-ph]];
P.~Colangelo and F.~De Fazio,
  JHEP {\bf 1806} (2018) 082
  [arXiv:1801.10468 [hep-ph]];
  F.~Feruglio, P.~Paradisi and O.~Sumensari,
  arXiv:1806.10155 [hep-ph];
  M.~A.~Ivanov, J.~G.~K\"orner and C.~T.~Tran,
  Phys.\ Rev.\ D {\bf 95} (2017) no.3,  036021
  [arXiv:1701.02937 [hep-ph]];
  A.~Azatov, D.~Bardhan, D.~Ghosh, F.~Sgarlata and E.~Venturini,
  arXiv:1805.03209 [hep-ph];
  Z.~R.~Huang, Y.~Li, C.~D.~Lu, M.~A.~Paracha and C.~Wang,
  arXiv:1808.03565 [hep-ph].


\bibitem{Gonzalez-Alonso:2017iyc} 
  M.~González-Alonso, J.~Martin Camalich and K.~Mimouni,
  Phys.\ Lett.\ B {\bf 772}, 777 (2017)
  [arXiv:1706.00410 [hep-ph]].




\bibitem{Li:2016vvp} 
  X.~Q.~Li, Y.~D.~Yang and X.~Zhang,
  JHEP {\bf 1608}, 054 (2016)
  [arXiv:1605.09308 [hep-ph]].


\bibitem{Alonso:2016oyd} 
  R.~Alonso, B.~Grinstein and J.~Martin Camalich,
  Phys.\ Rev.\ Lett.\  {\bf 118}, no. 8, 081802 (2017)
  [arXiv:1611.06676 [hep-ph]].


\bibitem{Dorsner:2017ufx} 
  I.~Doršner, S.~Fajfer, D.~A.~Faroughy and N.~Košnik,
  JHEP {\bf 1710}, 188 (2017)
  [arXiv:1706.07779 [hep-ph]].


\bibitem{Hiller:2014yaa} 
  G.~Hiller and M.~Schmaltz,
  Phys.\ Rev.\ D {\bf 90}, 054014 (2014)
  [arXiv:1408.1627 [hep-ph]];
  G.~Hiller and I.~Nisandzic,
  Phys.\ Rev.\ D {\bf 96}, no. 3, 035003 (2017)
  [arXiv:1704.05444 [hep-ph]];
  C.~Hati, G.~Kumar, J.~Orloff and A.~M.~Teixeira,
  arXiv:1806.10146 [hep-ph].


\bibitem{Becirevic:2017jtw} 
  D.~Bečirević and O.~Sumensari,
  JHEP {\bf 1708}, 104 (2017)
  [arXiv:1704.05835 [hep-ph]].

\bibitem{Assad:2017iib} 
  N.~Assad, B.~Fornal and B.~Grinstein,
  Phys.\ Lett.\ B {\bf 777}, 324 (2018)
  [arXiv:1708.06350 [hep-ph]].


\bibitem{Sakaki:2013bfa} 
  Y.~Sakaki, M.~Tanaka, A.~Tayduganov and R.~Watanabe,
  Phys.\ Rev.\ D {\bf 88}, no. 9, 094012 (2013)
  [arXiv:1309.0301 [hep-ph]].

\bibitem{Hiller:2016kry}
  G.~Hiller, D.~Loose and K.~Sch\"onwald,
  JHEP {\bf 1612} (2016) 027
  [arXiv:1609.08895 [hep-ph]].


\bibitem{Becirevic:2015asa}
  D.~Bečirević, S.~Fajfer and N.~Košnik,
  Phys.\ Rev.\ D {\bf 92} (2015) no.1,  014016
  [arXiv:1503.09024 [hep-ph]].
  
\bibitem{Cox:2016epl}
  P.~Cox, A.~Kusenko, O.~Sumensari and T.~T.~Yanagida,
  JHEP {\bf 1703} (2017) 035
  [arXiv:1612.03923 [hep-ph]].

\bibitem{Becirevic:2016yqi}
  D.~Bečirević, S.~Fajfer, N.~Košnik and O.~Sumensari,
  Phys.\ Rev.\ D {\bf 94} (2016) no.11,  115021
  [arXiv:1608.08501 [hep-ph]].



  
\bibitem{Bauer:2015knc} 
  M.~Bauer and M.~Neubert,
  Phys.\ Rev.\ Lett.\  {\bf 116}, no. 14, 141802 (2016)
  [arXiv:1511.01900 [hep-ph]].

\bibitem{Buttazzo:2017ixm} 
  D.~Buttazzo, A.~Greljo, G.~Isidori and D.~Marzocca,
  JHEP {\bf 1711}, 044 (2017)
  [arXiv:1706.07808 [hep-ph]].



\bibitem{DiLuzio:2017vat} 
  L.~Di Luzio, A.~Greljo and M.~Nardecchia,
  Phys.\ Rev.\ D {\bf 96}, no. 11, 115011 (2017)
  [arXiv:1708.08450 [hep-ph]];
  M.~Bordone, C.~Cornella, J.~Fuentes-Martin and G.~Isidori,
  Phys.\ Lett.\ B {\bf 779}, 317 (2018)
  [arXiv:1712.01368 [hep-ph]];
  L.~Calibbi, A.~Crivellin and T.~Li,
  arXiv:1709.00692 [hep-ph];
  M.~Blanke and A.~Crivellin,
  Phys.\ Rev.\ Lett.\  {\bf 121}, no. 1, 011801 (2018)
  [arXiv:1801.07256 [hep-ph]];
  R.~Barbieri and A.~Tesi,
  Eur.\ Phys.\ J.\ C {\bf 78}, no. 3, 193 (2018)
  [arXiv:1712.06844 [hep-ph]];
  A.~Greljo and B.~A.~Stefanek,
  Phys.\ Lett.\ B {\bf 782}, 131 (2018)
  [arXiv:1802.04274 [hep-ph]].


\bibitem{Fajfer:2015ycq} 
  S.~Fajfer and N.~Košnik,
  Phys.\ Lett.\ B {\bf 755}, 270 (2016)
  [arXiv:1511.06024 [hep-ph]].


\bibitem{Dorsner:2014axa} 
  I.~Dorsner, S.~Fajfer and A.~Greljo,
  JHEP {\bf 1410}, 154 (2014)
  [arXiv:1406.4831 [hep-ph]].

\bibitem{Sirunyan:2017yrk} 
  A.~M.~Sirunyan {\it et al.} [CMS Collaboration],
  JHEP {\bf 1707}, 121 (2017)
  [arXiv:1703.03995 [hep-ex]].


\bibitem{Sirunyan:2018nkj} 
  A.~M.~Sirunyan {\it et al.} [CMS Collaboration],
  arXiv:1803.02864 [hep-ex].


\bibitem{CMS:2018sgp} 
  CMS Collaboration [CMS Collaboration],
  CMS-PAS-EXO-17-003.


\bibitem{Diaz:2017lit} 
  B.~Diaz, M.~Schmaltz and Y.~M.~Zhong,
  JHEP {\bf 1710}, 097 (2017)
  [arXiv:1706.05033 [hep-ph]].


\bibitem{Camargo-Molina:2018cwu} 
  J.~E.~Camargo-Molina, A.~Celis and D.~A.~Faroughy,
  arXiv:1805.04917 [hep-ph].

\bibitem{CMS:2018itt} 
  CMS Collaboration [CMS Collaboration],
  CMS-PAS-B2G-16-027.


\bibitem{CMS:2018bhq} 
  CMS Collaboration [CMS Collaboration],
  CMS-PAS-SUS-18-001.


\bibitem{Blumlein:1996qp} 
  J.~Blumlein, E.~Boos and A.~Kryukov,
  Z.\ Phys.\ C {\bf 76}, 137 (1997)
  [hep-ph/9610408].



\bibitem{Faroughy:2016osc} 
  D.~A.~Faroughy, A.~Greljo and J.~F.~Kamenik,
  Phys.\ Lett.\ B {\bf 764}, 126 (2017)
  [arXiv:1609.07138 [hep-ph]].


\bibitem{Greljo:2017vvb} 
  A.~Greljo and D.~Marzocca,
  Eur.\ Phys.\ J.\ C {\bf 77}, no. 8, 548 (2017)
  [arXiv:1704.09015 [hep-ph]].






\bibitem{Aaboud:2017buh} 
  M.~Aaboud {\it et al.} [ATLAS Collaboration],
  JHEP {\bf 1710}, 182 (2017)
  [arXiv:1707.02424 [hep-ex]].


\bibitem{Aaboud:2017sjh} 
  M.~Aaboud {\it et al.} [ATLAS Collaboration],
  JHEP {\bf 1801}, 055 (2018)
  [arXiv:1709.07242 [hep-ex]].


\bibitem{Cowan:2010js} 
  G.~Cowan, K.~Cranmer, E.~Gross and O.~Vitells,
  Eur.\ Phys.\ J.\ C {\bf 71}, 1554 (2011)
  Erratum: [Eur.\ Phys.\ J.\ C {\bf 73}, 2501 (2013)]
  [arXiv:1007.1727 [physics.data-an]].


\bibitem{Alloul:2013bka} 
  A.~Alloul, N.~D.~Christensen, C.~Degrande, C.~Duhr and B.~Fuks,
  Comput.\ Phys.\ Commun.\  {\bf 185}, 2250 (2014)
  [arXiv:1310.1921 [hep-ph]].


\bibitem{Alwall:2014hca} 
  J.~Alwall {\it et al.},
  JHEP {\bf 1407}, 079 (2014)
  [arXiv:1405.0301 [hep-ph]].


\bibitem{Sjostrand:2014zea} 
  T.~Sjöstrand {\it et al.},
  Comput.\ Phys.\ Commun.\  {\bf 191}, 159 (2015)
  [arXiv:1410.3012 [hep-ph]].


\bibitem{deFavereau:2013fsa} 
  J.~de Favereau {\it et al.} [DELPHES 3 Collaboration],
  JHEP {\bf 1402}, 057 (2014)
  [arXiv:1307.6346 [hep-ex]].


\bibitem{Aaboud:2016hmk} 
  M.~Aaboud {\it et al.} [ATLAS Collaboration],
  Eur.\ Phys.\ J.\ C {\bf 76}, no. 10, 541 (2016)
  [arXiv:1607.08079 [hep-ex]];
  [arXiv:1807.06573 [hep-ex]].



\bibitem{Becirevic:2018afm} 
  D.~Bečirević, I.~Doršner, S.~Fajfer, D.~A.~Faroughy, N.~Košnik and O.~Sumensari,
  arXiv:1806.05689 [hep-ph].



\bibitem{Cai:2017wry} 
  Y.~Cai, J.~Gargalionis, M.~A.~Schmidt and R.~R.~Volkas,
  JHEP {\bf 1710}, 047 (2017)
  [arXiv:1704.05849 [hep-ph]].



\bibitem{Marzocca:2018wcf}
  D.~Marzocca,
  JHEP {\bf 1807} (2018) 121
  [arXiv:1803.10972 [hep-ph]].

\bibitem{Crivellin:2017zlb} 
  A.~Crivellin, D.~Müller and T.~Ota,
  JHEP {\bf 1709}, 040 (2017)
  [arXiv:1703.09226 [hep-ph]].


\bibitem{Patrignani:2016xqp} 
  C.~Patrignani {\it et al.} [Particle Data Group],
  Chin.\ Phys.\ C {\bf 40}, no. 10, 100001 (2016).

\bibitem{Miyazaki:2011xe}
  Y.~Miyazaki {\it et al.} [Belle Collaboration],
  Phys.\ Lett.\ B {\bf 699} (2011) 251
  [arXiv:1101.0755 [hep-ex]].


\bibitem{Feruglio:2016gvd} 
  F.~Feruglio, P.~Paradisi and A.~Pattori,
  Phys.\ Rev.\ Lett.\  {\bf 118}, no. 1, 011801 (2017)
  [arXiv:1606.00524 [hep-ph]];
  F.~Feruglio, P.~Paradisi and A.~Pattori,
  JHEP {\bf 1709}, 061 (2017)
  [arXiv:1705.00929 [hep-ph]];
  C.~Cornella, F.~Feruglio and P.~Paradisi,
  arXiv:1803.00945 [hep-ph].
 
 
  
\bibitem{Cirigliano:2007xi} 
  V.~Cirigliano and I.~Rosell,
  Phys.\ Rev.\ Lett.\  {\bf 99}, 231801 (2007)
  [arXiv:0707.3439 [hep-ph]].


\bibitem{Glattauer:2015teq} 
  R.~Glattauer {\it et al.} [Belle Collaboration],
  Phys.\ Rev.\ D {\bf 93}, no. 3, 032006 (2016)
  [arXiv:1510.03657 [hep-ex]].

\bibitem{Lees:2012zz}
  J.~P.~Lees {\it et al.} [BaBar Collaboration],
  Phys.\ Rev.\ D {\bf 86} (2012) 012004
  [arXiv:1204.2852 [hep-ex]].
  
\bibitem{DiLuzio:2017chi}
  L.~Di Luzio and M.~Nardecchia,
  Eur.\ Phys.\ J.\ C {\bf 77} (2017) no.8,  536
  [arXiv:1706.01868 [hep-ph]].

\bibitem{Bordone:2018nbg}
  M.~Bordone, C.~Cornella, J.~Fuentes-Mart\'in and G.~Isidori,
  arXiv:1805.09328 [hep-ph];
  L.~Di Luzio, J.~Fuentes-Martin, A.~Greljo, M.~Nardecchia and S.~Renner,
  arXiv:1808.00942 [hep-ph].

    
\end{thebibliography}
\end{document}